\documentclass[aps,prb,twocolumn,floatfix]{revtex4-1}
\usepackage{amsmath,amssymb,bm,epsfig,graphicx,longtable}
\usepackage[caption=false]{subfig} \usepackage{comment} \usepackage{siunitx}

\begin{document}

\title{Limitations of the DFT--1/2 method for covalent semiconductors and
transition-metal oxides}

\author{Jan Doumont} \author{Fabien Tran} \author{Peter Blaha}
\affiliation{Institute of Materials Chemistry, Vienna University of
Technology, Getreidemarkt 9/165-TC, A-1060 Vienna, Austria}

\date{\today}

\begin{abstract}

The DFT--1/2 method in density functional theory [L. G. Ferreira \textit{et
al}., Phys. Rev. B \textbf{78}, 125116 (2008)] aims to provide accurate band
gaps at the computational cost of semilocal calculations. The method has
shown promise in a large number of cases, however some of its limitations or
ambiguities on how to apply it to covalent semiconductors have been pointed
out recently [K.-H. Xue \textit{et al}., Comput. Mater. Science
\textbf{153}, 493 (2018)]. In this work, we investigate in detail some of
the problems of the DFT--1/2 method with a focus on two classes of
materials: covalently bonded semiconductors and transition-metal oxides. We
argue for caution in the application of DFT--1/2 to these materials, and the
condition to get an improved band gap is a spatial separation of the
orbitals at the valence band maximum and conduction band minimum.

\end{abstract}

\maketitle

\section{Introduction}
\label{sec:introduction}

The calculation of the fundamental band gap of solids in Kohn-Sham (KS)
density functional theory\cite{hohenbergInhomogeneousElectronGas1964a,
  kohnSelfConsistentEquationsIncluding1965} (DFT) is a long standing
problem.\cite{perdewDensityFunctionalTheory2009} The reason is that the
exchange--correlation functional of the local density
approximation\cite{kohnSelfConsistentEquationsIncluding1965} (LDA) severely
underestimates band gaps by typically 50 --
100\%,\cite{perdewDensityFunctionalTheory2009} and the standard functionals
of the generalized gradient approximation (GGA)\cite{PerdewPRL96} do not
perform much better.\cite{HeydJCP05} The current state-of-the-art in band
gap calculations is Hedin's $GW$
method,\cite{aryasetiawanGWMethod1998,hedinCorrelationEffectsElectron1999}
but it goes beyond DFT and is computationally very demanding especially if
applied self-consistently.\cite{ShishkinPRL07} Within the generalized
Kohn--Sham (gKS) scheme\cite{seidlGeneralizedKohnShamSchemes1996} (i.e.,
with non-multiplicative potentials), hybrid functionals, which mix LDA/GGA
functionals with exact exchange,\cite{BeckeJCP93b} do offer greatly improved
band gaps,\cite{HeydJCP05} but at a computational cost that is also much
higher (by one or two orders of magnitude) than LDA/GGA functionals. The
meta-GGA (MGGA) approximation,\cite{DellaSalaIJQC16} which is also of the
semilocal type and therefore computationally fast, is a very promising
route for improving band gaps within the gKS framework at a modest cost.
The MGGA functionals that have been developed so far are however not as
accurate as the hybrid or $GW$ methods.\cite{XiaoPRB13,YangPRB16,JanaJCP18}

Nevertheless, within the true KS--DFT scheme, i.e., with a multiplicative
potential, computationally fast DFT methods have been developed for band gap
calculations, like the functional of Armiento and
K\"ummel,\cite{ArmientoPRL13,VlcekPRB15} the potential of Gritsenko
\textit{et al}.\cite{GritsenkoPRA95,KuismaPRB10} (GLLB), or the modified
Becke-Johnson potential\cite{tranAccurateBandGaps2009} (mBJ), the latter
being as accurate as the very expensive hybrid or $GW$ methods.

Another fast method designed for band gaps is
DFT--1/2,\cite{ferreiraApproximationDensityFunctional2008} which is an
application of Slater's half-occupation (transition state)
technique\cite{slaterStatisticalExchangeCorrelationSelfConsistent1972,
  slaterSelfConsistentFieldEnsuremathAlpha1972} to periodic solids. It only
requires the addition of a self-energy correction potential,
calculated from a half-ionized free atom, to the usual KS--DFT
potential (see Sec.~\ref{sec:theory} for details). The method has been
shown to perform quite well for a number of test
sets,\cite{ferreiraLDA1TechniqueRecent2013,
  ferreiraSlaterHalfoccupationTechnique2011,
  rodriguespelaLDA1MethodImplemented2017} and has been evaluated as a good
starting point for $G_{0}W_{0}$
calculations.\cite{rodriguespelaProbingLDA1Method2016} For instance, an
application to metal halide perovskites has found comparable accuracy to
$GW$.\cite{taoAccurateEfficientBand2017} Thanks to its low computational
cost, DFT--1/2 has been regularly applied to systems that require larger
unit cells. A study of the negatively charged nitrogen-vacancy center in
diamond has been performed with a generalized version of DFT--1/2, which is
suited not only for band gap but also optical transitions and defect
levels.\cite{lucattoGeneralProcedureCalculation2017} Other applications
include studies of doped
materials,\cite{ribeiroApplicationGGA1Excitedstate2015,
  belabbesMagnetismClusteringCrdoped2010}
heterostructures,\cite{santosDigitalMagneticHeterostructures2012,
  ribeiroInitioCalculationCdSe2012}
surfaces,\cite{belabbesElectronicPropertiesPolar2011,
  kufnerEnergeticsApproximateQuasiparticle2012a} or
interfaces.\cite{ribeiroFirstprinciplesCalculationAlAs2011,
  ribeiroAccuratePredictionSi2009} Also, in a study of semiconducting indium
alloys comparing the DFT--1/2 method with hybrid functionals, it was found
that, although the hybrid functionals were slightly more accurate, DFT--1/2
allows for larger supercells and consequently better convergence of the
bowing parameter.\cite{pelaComparingLDA1HSE032015} Another
comparative study of DFT--1/2 to the pseudo--self-interaction--corrected
approach to DFT was performed on
fluorides.\cite{matusalemElectronicPropertiesFluorides2018} Furthermore, a
few magnetic systems have been studied, namely
GaMnAs\cite{pelaGaMnAsPositionMn2012} and InN doped with
Cr.\cite{belabbesMagnetismClusteringCrdoped2010} We also mention that the
method has recently been applied successfully for the calculation of the
ionization potential of atoms and
molecules.\cite{rodriguespelaLDA1MethodApplied2018}

However, the limitations of the method have not been given much
consideration until recently.\cite{xueImprovedLDA1Method2018} These
limitations stem from the fact that the correction applied in DFT--1/2 has
an atomic origin. One of them is the application of the method to covalently
bonded semiconductors. Originally, it was argued that group IV
semiconductors (diamond, Si, and Ge) need a modified correction that is
calculated from a 1/4-ionized atom instead of a 1/2-ionized atom, the
argument being that valence band holes of neighboring atoms
overlap.\cite{ferreiraApproximationDensityFunctional2008} In III--V
compounds (GaAs, AlP, \ldots) it is claimed that the valence band hole
resembles more closely the photoionization hole in the atom, such that the
standard 1/2-ionization is
justified.\cite{ferreiraApproximationDensityFunctional2008}

As shown and discussed in detail in this work, another limitation of the
DFT--1/2 method is that it performs very poorly for transition-metal (TM)
oxides. Many of these materials are Mott insulators, where both the highest
occupied band and the lowest conduction band have strong TM $d$--orbital
characters which differ only by their angular shape, such that the spherical
atomic DFT--1/2 correction can not work efficiently for the band gap.

The focus of the present work will be on the problems of the DFT--1/2 method
mentioned above, namely the ambiguity about the ionization of the free atom
to calculate the correction potential in semiconductors and the limited
applicability of DFT--1/2 for TM oxides.

The paper is organized as follows. Section~\ref{sec:theory} provides a
description of the methods and the computational details, while the results
are presented and discussed in Sec.~\ref{sec:results}. Finally,
Sec.~\ref{sec:summary} gives the summary of this work.

\section{Theory}
\label{sec:theory}

In KS--DFT, the so-called KS band gap $E_{g}^{\mathrm{KS}}$ is defined as
the difference between the KS eigenvalues of the highest occupied
[$\epsilon_{N}(N)$] and lowest unoccupied [$\epsilon_{N+1}(N)$] orbitals of
the $N$--electron system. On the other hand, the fundamental band gap
$E_{g}$, the physical many-body property one is interested in, is defined as
the ionization potential $I(N)$ minus the electron affinity $A(N)$ and can
be expressed in terms of the (exact) KS eigenvalues of the highest occupied
orbitals of the $N$-- and $N+1$--electron
systems:\cite{perdewDensityFunctionalTheory2009,
  perdewPhysicalContentExact1983}
\begin{align} E_g & = I \left( N \right) - A \left( N \right)& \nonumber \\
& = - \epsilon_N \left( N \right) - \left( - \epsilon_{N+1} \left( N+1
\right) \right) & \nonumber \\ & = \underbrace{\epsilon_{N+1}\left( N
\right) - \epsilon_N \left( N \right)}_{E_{g}^{\mathrm{KS}}} +
\underbrace{\epsilon_{N+1} \left( N + 1 \right) - \epsilon_{N+1} \left( N
\right)}_{\Delta_{\mathrm{xc}}} &\nonumber \\
\label{eq:4} & = E_{g}^{\mathrm{KS}} + \Delta_{\mathrm{xc}},&
\end{align} where $\Delta_{\mathrm{xc}}$ is the discontinuity of the
exchange--correlation potential at integer values of the number of electrons
$N$. In KS--DFT calculations employing LDA or GGA functionals, this
discontinuity is not captured\cite{yangDerivativeDiscontinuityBandgap2012}
(but can be calculated by some means in finite
systems\cite{AndradePRL11,ChaiPRL13,KraislerJCP14}).  From Eq.~(\ref{eq:4}),
it is clear that a good estimation of the true band gap can, in principle,
not be obtained by considering $E_{g}^{\mathrm{KS}}$ alone in particular
since $\Delta_{\mathrm{xc}}$ can be of the same order of magnitude as the
band gap itself.\cite{GruningJCP06,GruningPRB06}

The DFT--1/2 technique aims to correct the band gap problem by adapting
Slater's atomic transition state technique to periodic solids. Starting from
Janak's theorem,\cite{janakProofThatFrac1978}
\begin{equation}
\label{eq:5} \frac{\partial E(f_{\alpha})}{\partial f_{\alpha}} =
\epsilon_{\alpha} \left( f_{\alpha} \right),
\end{equation} where $E(f_{\alpha})$ is the total energy of the system and
$f_{\alpha}$ is the occupation number of orbital $\phi_{\alpha}$ relative to
the neutral atom ($f_{\alpha}=0$), and using the midpoint rule for
integrating the right-hand side of Eq.~(\ref{eq:5}), it is trivial to show
that the KS eigenvalue for the 1/2-ionized (hence transition) state can be
used to calculate the ionization potential of the atom:
\begin{equation}
\label{eq:6} E \left( 0 \right) - E \left( -1 \right) \simeq
\epsilon_{\alpha} \left( -1/2 \right).
\end{equation}

In order to benefit from Eq.~(\ref{eq:6}) for self-consistent DFT
calculations in solids,\cite{ferreiraApproximationDensityFunctional2008,
  ferreiraSlaterHalfoccupationTechnique2011} a self-energy correction
potential $v_S$ is defined by rewriting the ionization potential the
following way:
\begin{align} E \left( 0 \right) - E \left( -1 \right) = \epsilon_{\alpha}
\left( 0 \right) - \int \mathrm{d}^3 \vec{r} \rho_{\alpha} \left( \vec{r}
\right) v_S \left( \vec{r} \right),
  \label{eq:7} \intertext{with $\rho_{\alpha} =
\left\vert\phi_{\alpha}\right\vert^{2}$ and where $v_S$ is chosen such that}
  \label{eq:8} \int\mathrm{d}^3 \vec{r} \; \rho_{\alpha} \left( \vec{r}
\right) v_S \left( \vec{r} \right) \simeq \epsilon_{\alpha} \left( 0\right)
- \epsilon_{\alpha} \left( -1/2 \right)
\end{align} and therefore Eq.~(\ref{eq:6}) satisfied. Equation~(\ref{eq:8})
shows that adding $-v_S$ to the effective KS potential $v_{\mathrm{KS}}$ in
a calculation should shift the eigenvalue of orbital $\alpha$ by
$\epsilon_{\alpha} \left( -1/2 \right) - \epsilon_{\alpha} \left(0\right)$
and therefore bring it close to $\epsilon_{\alpha} \left( -1/2 \right)$,
i.e., the ionization potential according to Eq.~(\ref{eq:6}).  In practice,
the potential $v_S$ is not obtained from calculations on the solid, but on
an isolated atom (the one where the orbital $\alpha$ is mostly
located):\cite{ferreiraApproximationDensityFunctional2008}
\begin{equation}
\label{eq:9} v_S = v_{\mathrm{KS}}^{\mathrm{atom}} \left( f_{\alpha} = 0
\right) - v_{\mathrm{KS}}^{\mathrm{atom}} \left(f_{\alpha} = -1/2 \right),
\end{equation} where $v_{\mathrm{KS}}^{\mathrm{atom}}$ are the KS effective
potentials obtained at the end of self-consistent calculations in the
neutral and 1/2-ionized states.

Concretely, the DFT--1/2 method consists, first, of two self-consistent
calculations on the free atom to calculate $v_{S}$ with Eq.~(\ref{eq:9}),
and the orbital $\phi_{\alpha}$ that is chosen to be ionized is the one that
is supposed to contribute the most to the valence band maximum (VBM) in the
solid. Then, this atomic potential $v_{S}$ is added to the usual LDA or GGA
effective KS potential $v_{\mathrm{KS}}$ for the self-consistent calculation
on the solid. However, before $v_S$ is added to $v_{\mathrm{KS}}$, it must be
multiplied by a spherical step function
\begin{align} \Theta \left( r \right) & =
                                             \begin{cases} \left( 1 - \left(
\frac{r}{r_{c}} \right)^8 \right)^3 \qquad & r \leq r_{c} \\ 0 \qquad & r >
r_{c}
                                             \end{cases}\label{eq:11}
\end{align} because $v_S$ falls off only like $1/r$ at long range which
causes divergence when summed over the lattice. The cutoff radius $r_{c}$ is
the only parameter introduced in the method, and is determined variationally
by maximizing the band gap.\cite{ferreiraApproximationDensityFunctional2008}

As argued in Ref.~\onlinecite{ferreiraSlaterHalfoccupationTechnique2011},
the correction to the KS band gap due to $v_S$ can be somehow identified to
the discontinuity $\Delta_{\text{xc}}$ in Eq.~(\ref{eq:4}) (although it is
questionable since the potential is still
multiplicative\cite{KuemmelRMP08,yangDerivativeDiscontinuityBandgap2012}).
However, we mention that no correction was applied to the conduction band
minimum (CBM). As reported in
Ref.~\onlinecite{ferreiraSlaterHalfoccupationTechnique2011}, such correction
should affect only little the unoccupied states due to their more
delocalized nature.

A few extensions or refinements to the method have been proposed. The shell
correction from Ref.~\onlinecite{xueImprovedLDA1Method2018} uses a
step-function with an additional (inner) radius to improve the accuracy and
will be discussed in detail in Sec.~\ref{sec:shell-correction-lda}. In other
works,\cite{ataideFastAccurateApproximate2017,
  ribeiroInitioQuasiparticleApproximation2015} an empirical amplification
factor (which multiplies $v_S$ by a constant) to fit experiment was used. In
Ref.~\onlinecite{ataideFastAccurateApproximate2017}, non-standard ionization
levels for the correction potential (other than 1/4 or 1/2) have been
used. The character of the atomic orbital contributions to the VBM is used
to determine the ionization levels (normalized to 1/2 across both atomic
species). In Ref.~\onlinecite{lucattoGeneralProcedureCalculation2017} a
generalization of DFT--1/2 also suited for optical transition levels
(including adding self-energy correction to the excited band, and
non-standard ionization levels) has been applied to the NV$^-$ center of
diamond.

For the present work, the DFT--1/2 method has been implemented into the
all-electron \textsc{wien2k}\cite{wien2k} code which is based on the
linearized-augmented plane-wave (LAPW) method.\cite{AndersenPRB75,Singh} The
implementation is very similar to the one reported
recently\cite{rodriguespelaLDA1MethodImplemented2017} in \textsc{exciting}
which is also an LAPW-based code.  The calculations were done at the
experimental lattice parameters (specified in Table~S1 of
Ref.~\onlinecite{SM_LDA-half}) for all compounds. A dense
$24\times24\times24$ $\mathbf{k}$--mesh was used for all cubic solids, while
for other structures a proportional mesh with 24 $\mathbf{k}$--points along
the direction corresponding to the shortest lattice constant was used. For
some of the TM oxides [notably those with antiferromagnetic (AFM) ordering,
which have larger unit cells], a less dense $\mathbf{k}$--mesh was used, but
care was taken that convergence is reached. The same applies to the basis
set size. For all compounds containing Ga or heavier atoms, the calculations
were done with spin-orbit coupling included. The cutoff radius $r_{c}$ in
Eq.~(\ref{eq:11}) is optimized using a multi-dimensional search with a
precision of \SI{0.01}{\electronvolt}, which corresponds to a precision in
$r_{c}$ of about \SI{0.05}{\bohr} (the band gap is not very sensitive to
$r_{c}$ close to the extremum). Furthermore, the optimal cutoff radii of
different atoms in binary compounds are to a large extent
independent.\cite{ferreiraApproximationDensityFunctional2008} LDA and GGA
[using the functional of Perdew \textit{et al}. \cite{PerdewPRL96} (PBE)]
calculations were done with and without the 1/2 correction. For comparisons
purpose, calculations with the mBJ
potential,\cite{tranAccurateBandGaps2009,TranJPCA17,TranPRM18} which has
been shown to be the most accurate semilocal potential for band gap
calculations and is even superior to hybrid
functionals,\cite{LeePRB16,TranJPCA17,NakanoJAP18,TranPRM18} will also be
reported.

\section{Results}
\label{sec:results}

\subsection{Group IV and III-V semiconductors}
\label{sec:group-iv-se}

\begin{table*}[htb]
\caption{Band gaps (in eV) of groups IV and III-V semiconductors calculated
using the DFT--1/2 method with different underlying functionals (LDA or PBE)
and ionization degrees (1/2 or 1/4). The orbitals that were ionized and the
cutoff radii $r_{c}$ (in $a_{0}$) in Eq.~(\ref{eq:11}) are indicated in the
second and third columns, respectively. Only the cutoff radii from LDA--1/2
calculations are shown, however we checked that those for PBE--1/2,
LDA--1/4, and PBE--1/4 calculations are practically identical (the
difference is below \SI{0.05}{\bohr}, which does not impact the band gap). A
dash ``-'' in the orbitals column indicates a non-corrected atom. Literature
results (all based on LDA) are also given and are marked according to the
reference/code: (S) for \textsc{siesta} and (V) for \textsc{vasp}
calculations,\cite{ferreiraApproximationDensityFunctional2008} (E) for
\textsc{exciting} calculations,\cite{rodriguespelaLDA1MethodImplemented2017}
and (X) for \textsc{vasp} calculations.\cite{xueImprovedLDA1Method2018} When
the correction details of the literature results do not correspond to ours (which
atoms, orbitals are corrected), they are specified in parenthesis. Note that
in Ref.~\onlinecite{xueImprovedLDA1Method2018}, the calculations were done
at the LDA lattice constant. For comparison purposes, LDA, PBE, and mBJ
results are also shown. The experimental results are from
Refs.~\onlinecite{luceroImprovedSemiconductorLattice2012} and
\onlinecite{crowleyResolutionBandGap2016a}. The MAE (in eV) and MARE (in \%)
are calculated when the correction is applied on all atoms and for an
ionized $p$--orbital in Ga (as deduced from a partial charge analysis at
VBM). The most accurate values among the DFT--1/2 methods are underlined.}
\begin{ruledtabular}
\begin{tabular}{lccccccccccc} Solid & Orbitals & $r_{c}$ & LDA & LDA--1/4 &
LDA--1/2 & PBE & PBE--1/4 & PBE--1/2 & mBJ & Other works & Expt. \\ \hline C
& $p$ & 2.41 & 4.10 & 4.95 & \underline{5.82} & 4.14 & 5.04 & 5.95 & 4.92 &
5.25 (S,1/4-$sp$) & 5.50 \\ Si & $p$ & 3.78 & 0.47 & \underline{1.20} & 1.96
& 0.57 & 1.35 & 2.16 & 1.15 & 1.21 (S,1/4) & 1.17 \\ SiC & $p$, $p$ & 2.80,
3.00 & 1.31 & 2.31 & 3.40 & 1.35 & \underline{2.43} & 3.59 & 2.25 & 2.32
(E)\footnotemark[1] & 2.42 \\ SiC & -, $p$ & 3.00 & 1.31 & 2.19 & 3.16 &
1.35 & \underline{2.28} & 3.31 & 2.25 & & 2.42 \\ Ge & $p$ & 4.11 & metal &
0.04 & 0.37 & metal & 0.27 & \underline{0.59} & 0.76 & 0.27 (X,1/4) & 0.74
\\ BN & $p$, $p$ & 2.41, 2.47 & 4.35 & 5.24 & \underline{6.78} & 4.47 & 5.79
& 7.06 & 5.80 & & 6.36 \\ BP & $p$, $p$ & 3.17, 3.09 & 1.18 &
\underline{1.97} & 2.79 & 1.24 & 2.48 & 2.95 & 1.85 & & 2.10 \\ BAs & $p$,
$p$ & 3.22, 3.25 & 1.04 & \underline{1.82} & 2.62 & 1.09 & 1.93 & 2.77 &
1.58 & & 1.46 \\ AlN & $p$, $p$ & 2.91, 2.92 & 3.25 & 4.48 & 5.80 & 3.34 &
\underline{4.66} & 6.08 & 4.88 & & 4.90 \\ AlP & $p$, $p$ & 3.74, 3.69 &
1.45 & 2.29 & 3.15 & 1.59 & \underline{2.50} & 3.43 & 2.31 & 2.23 (X,1/4) &
2.5 \\ AlP & -, $p$ & 3.69 & 1.45 & 2.18 & 2.95 & 1.59 & \underline{2.38} &
3.21 & 2.31 & 2.96 (E)\footnotemark[1] & 2.5 \\ AlAs & $p$, $p$ & 4.39, 3.88
& 1.25 & 2.08 & 2.92 & 1.35 & \underline{2.26} & 3.17 & 2.05 & 2.73
(V,1/2-As) & 2.23 \\ AlSb & $p$, $p$ & 6.28, 4.11 & 0.89 & 1.51 & 2.17 &
0.97 & \underline{1.62} & 2.30 & 1.51 & 1.97 (X,1/2-Sb) & 1.69 \\ GaN & $p$,
$p$ & 1.30, 3.00 & 1.66 & 2.50 & \underline{3.42} & 1.66 & 2.55 & 3.53 &
2.86 & & 3.28 \\ GaN & $d$, $p$ & 1.30, 3.00 & 1.66 & 2.56 &
\underline{3.54} & 1.66 & 2.61 & 3.66 & 2.86 & 3.56 (E),\footnotemark[1]
3.52 (V,1/2) & 3.28 \\ GaP & $p$, $p$ & 1.15, 3.50 & 1.41 & 2.00 &
\underline{2.51} & 1.57 & \underline{2.21} & 2.80 & 2.22 & 2.57 (X,1/2-P) &
2.35 \\ GaAs & $p$, $p$ & 1.14, 3.82 & 0.19 & 0.71 & 1.27 & 0.43 & 0.97 &
\underline{1.57} & 1.54 & 1.41 (V,1/2-As) & 1.52 \\ GaAs\footnotemark[2] &
$d$, $p$ & 1.15, 3.83& 0.30 & 0.86 & \underline{1.47} & 0.54 & 1.13 & 1.77
& 1.64 & 1.46 (E)\footnotemark[1] & 1.52 \\ GaSb & $p$, $p$ & 1.14, 4.19 &
metal & 0.05 & 0.48 & metal & 0.30 & \underline{0.75} & 0.74 & 0.67
(X,1/2-Sb) & 0.82 \\ \hline MAE & & & 1.10 & 0.44 & 0.56 & 1.02 & 0.32 &
0.67 & 0.20 & & \\ MARE & & & 52 & 25 & 30 & 48 & 19 & 32 & 7 & & \\
\end{tabular}
\end{ruledtabular} \footnotetext[1]{Reference
\onlinecite{rodriguespelaLDA1MethodImplemented2017} does not provide details
about the ionization correction, but the very close agreement with one of
our results indicates which ionization correction was applied.}
\footnotetext[2]{Calculation without spin-orbit coupling for comparison with
the result from other work. The effect of the spin-orbit coupling is to
reduce the band gap by about \SI{0.1}{\electronvolt}.}
  \label{tab:g4-gaps}
\end{table*}

\begin{figure*}[htbp!]
\subfloat[]{\includegraphics[width=.99\columnwidth]{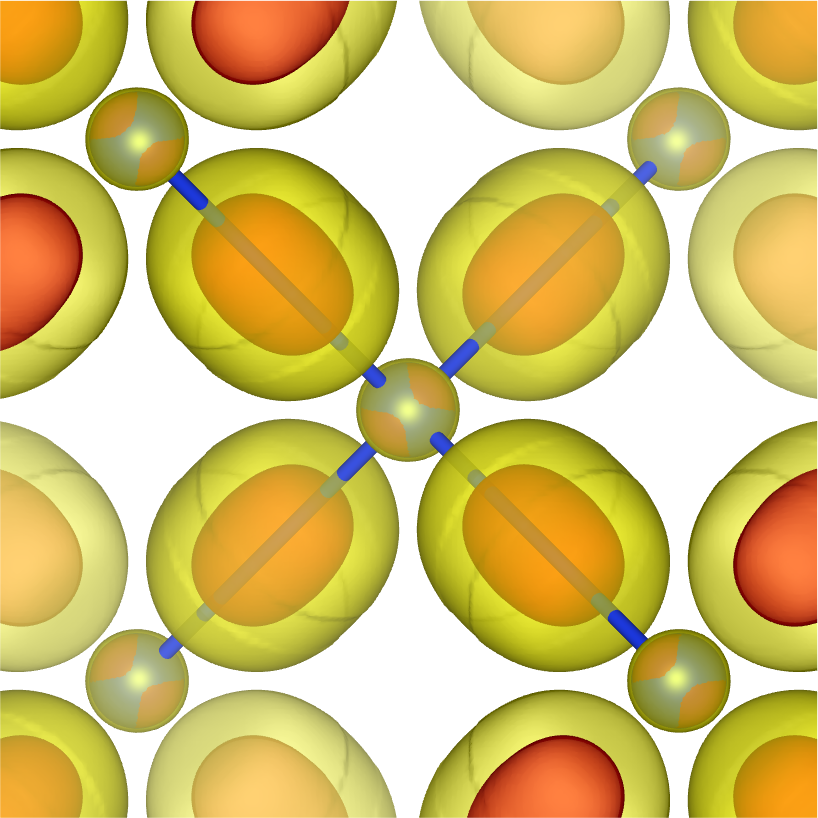}}
  \label{fig:Si-val-dens}\hspace{\columnsep}
\subfloat[]{\includegraphics[width=.99\columnwidth]{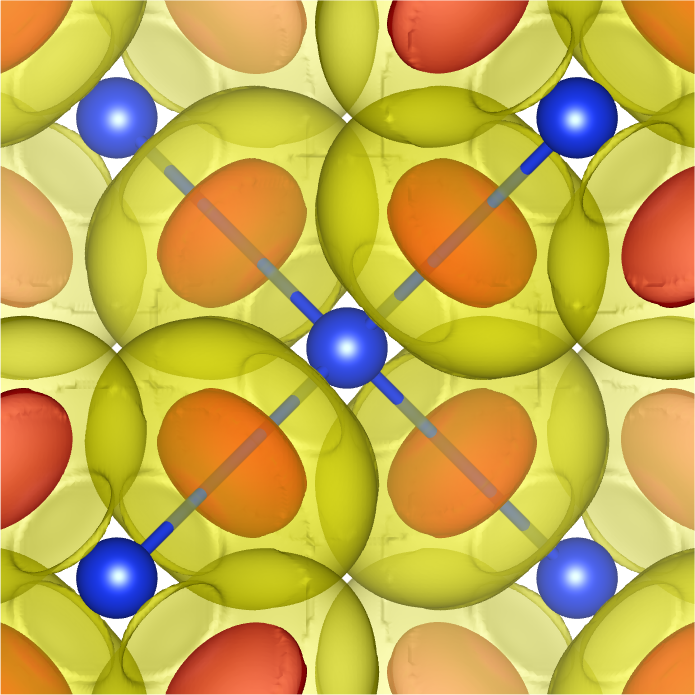}}
  \label{fig:Si-VS}
\caption{Illustration of the ``overlapping holes'' argument in
silicon. Isosurfaces are chosen for best visibility, in both cases with a
lower value in yellow and a higher one in red. The electron density close to
the VBM (a) sits mostly around the bond centers. The correction potential
$v_S$ (using PBE--1/2) (b) is the largest in the middle of the Si--Si bonds,
since the sum of the cutoff radii $r_{c}$ is larger than the
nearest-neighbor distance, even though the $v_S$ of the individual atoms is
spherically distributed.}
\label{fig:g4-overlapping-holes}
\end{figure*}

We start by mentioning that how to apply the DFT--1/2 method to the group IV
semiconductors C, Si, and Ge is unclear. In contrast to binary compounds,
the self-energy correction potential $v_{S}$ has not always been calculated
from 1/2-ionized free atoms, but from 1/4-ionized ones [i.e., with
$f_{\alpha}=-1/4$ in the second term of Eq.~(\ref{eq:9})]. Actually, for
diamond $v_S$ was calculated in
Refs.~\onlinecite{ferreiraApproximationDensityFunctional2008,
ferreiraSlaterHalfoccupationTechnique2011} by ionizing both the $p$-- and
$s$--bands by a 1/4-electron charge (in total, removing half an electron),
whereas for Si and Ge only the $p$--band receives a 1/4-ionization
correction. The argument behind this is that the orbital at the VBM overlaps
with the correction potential of both atoms in the unit cell, such that only
a 1/4 electron should be removed on each atom to avoid a correction that is
too large. This is illustrated for Si in
Fig.~\ref{fig:g4-overlapping-holes}, where we can see that $v_{S}$ is the
largest at the Si--Si bond center.

Turning to our DFT--1/2 calculations, Table~\ref{tab:g4-gaps} shows the
results for a set of covalent semiconductors that were obtained with a 1/2-
or 1/4-ionization correction. Furthermore, both LDA and PBE were considered
for the underlying semilocal functional. All atoms were corrected and the
ionized orbital is the one with the largest contribution to the VBM. For SiC
and AlP, an additional calculation was done where the correction is applied
only to the anion.

Indeed, we can see that the band gaps obtained using a 1/4-ionization
correction (i.e., LDA--1/4 and PBE--1/4) are very accurate for Si and SiC,
since the values differ by at most \SI{0.2}{\electronvolt} compared to
experiment, while using a 1/2-ionization correction (i.e., LDA--1/2 and
PBE--1/2) leads to overestimations of at least \SI{0.8}{\electronvolt}. For
diamond, the results show that using a 1/2-ionized (1/4-ionized) correction
leads to an overestimation (underestimation) of about
\SI{0.5}{\electronvolt}. For Ge, the experimental gap of
\SI{0.74}{\electronvolt} lies above the LDA--1/2 and PBE--1/2 values by
about 0.4 and \SI{0.2}{\electronvolt}, respectively, while using a
1/4-ionized correction leads to strongly underestimated values. Note the
contrast between Si and Ge which require different ionization, despite
having relatively similar valence band density and optimized cutoff radius
$r_{c}$ in Eq.~(\ref{eq:11}).

Another issue that may arise is the ambiguity in choosing the atom(s) and/or
orbital(s) on which the correction should be applied. For instance in the
case of binary semiconductors, it has been
claimed\cite{ferreiraApproximationDensityFunctional2008} that in most cases
(but not always) only the correction on the anion has an impact on the
results. While this may be true for ionic solids, where the states at the
VBM come only from the anion, such choice can not be always justified in the
case of binary semiconductors where both atoms may contribute to the
VBM. Thus, in addition to the degree of ionization correction (e.g., 1/4 or
1/2), it may not be always clear on which atoms the potential $v_{S}$ should
be applied.  Since in SiC the VBM has a dominant $p$--orbital character from
the C atom, we did an alternative calculation where the correction is
applied only to the C atom. Compared to the usual procedure where the
orbitals on all atoms are corrected, a reduction of the band gap by
\SIrange{0.1}{0.3}{\electronvolt} is observed. Good agreement with
experiment is obtained with 1/4-ionized correction (even though there is
very little correction potential overlap at the VBM in this case), while a
1/2-ionized correction leads to large overestimations of
\SI{\sim1}{\electronvolt} similar to Si.

Considering the III--V compounds, we see that LDA--1/2 and PBE--1/2 clearly
overestimate the band gaps for the B$X$ and Al$X$ compounds, while a
moderate overestimation is observed for GaN and GaP. On the other hand,
PBE--1/2 performs very well for GaAs and GaSb since the error is below
\SI{0.1}{\electronvolt}.

On average, PBE--1/4 is the most accurate of the DFT--1/2 methods for this
test set. It provides in eight cases the best agreement with experiment and
leads to a MAE of only \SI{0.32}{\electronvolt}; this is half of the one for
PBE--1/2 (\SI{0.67}{\electronvolt}) which is the worst of the DFT--1/2
methods. However, note that the mBJ potential which has MAE of
\SI{0.20}{\electronvolt} and MARE of 7\% is clearly more accurate. In
comparison, LDA and PBE lead to MAE that are around
\SI{1}{\electronvolt}. The general observation is that a 1/4 ionization is
more appropriate for the light systems, but not sufficient for the heavier
ones, i.e., those with Ga or Ge atoms, for which a 1/2-ionization
correction, either with LDA or PBE, is usually more suitable. Nevertheless,
a few borderline cases are C, BN, and GaP, where the best correction also
depends on the underlying semilocal functional. We also mention that for
only one system (GaN), there is no overlap (loosely defined as whether the
sum of the cutoff radii of two nearest-neighboring atoms is larger than
their distance) between the correction potentials $v_{S}$, while for the
other Ga compounds the overlap is small (tenths of one $a_{0}$, compared to
an overlap of \SIrange{3}{4}{\bohr} in Si and Ge).

\begin{figure}[htbp!]
\includegraphics[width=\columnwidth]{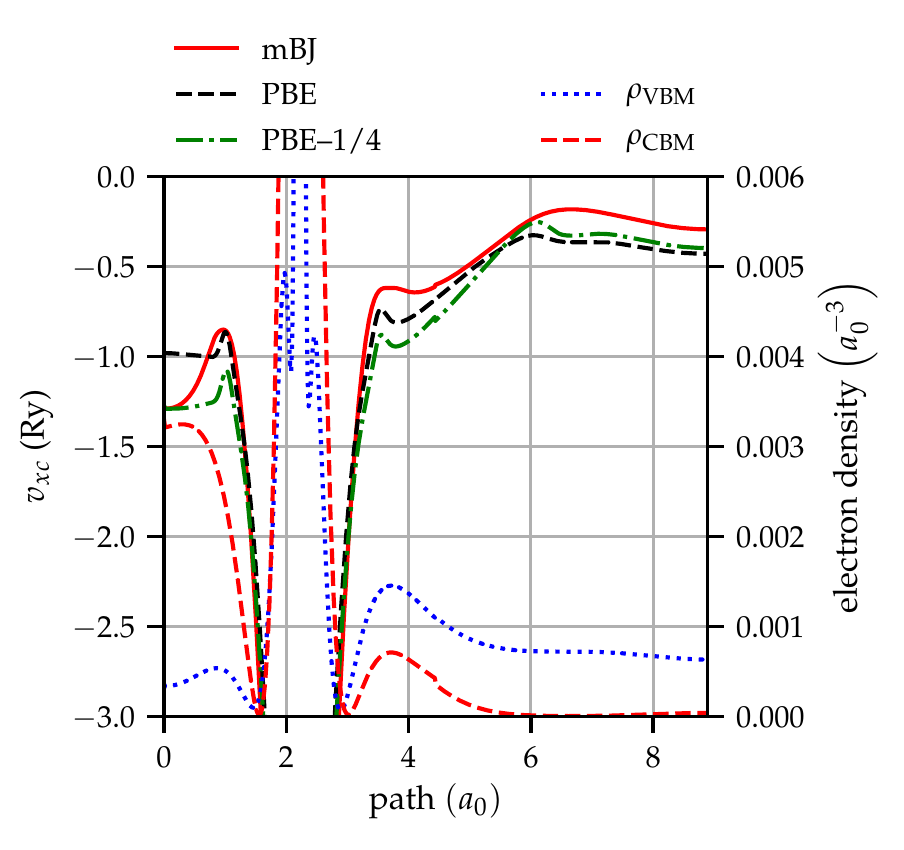}
\caption{Plots of mBJ, PBE, and PBE--1/4 exchange--correlation potentials
$v_{xc}$ in Si and densities of the VBM and CBM. The path is from (0, 0, 0)
to (1/2, 1/2, 1/2) in the unit cell fractional coordinates, thus starting at
the midpoint of the Si--Si bond, passing through the atom at (1/8, 1/8, 1/8)
and terminating at the middle of the unit cell. The densities are taken from
the PBE calculation.}
  \label{fig:Si-vxc}
\end{figure}

In Fig.~\ref{fig:Si-vxc}, the exchange--correlation potentials $v_{xc}$ mBJ,
PBE, and PBE--1/4 in Si are compared. The band gaps from mBJ
(\SI{1.15}{\electronvolt}) and PBE--1/4 (\SI{1.35}{\electronvolt}) are
relatively close to each other, but the corresponding potentials show
noticeable differences.
Compared to PBE, PBE--1/4 corrects the band gap by lowering the
energy in the region where the VBM density $\rho_{\text{VBM}}$ is very large
(in the region within \SI{2}{\bohr} from the atom), whereas mBJ has a
smaller correction. On the other hand, at the CBM mBJ leads to a larger
up-shift than PBE--1/4.

Concerning the orbital to which the ionization should be applied, the
Ga compounds are interesting since they are not always treated the
same way. For some reported
calculations,\cite{ferreiraApproximationDensityFunctional2008,
  pelaGaMnAsPositionMn2012} the $d$--orbital was ionized for all Ga
compounds, while in
Ref.~\onlinecite{ribeiroFirstprinciplesCalculationAlAs2011}, the Ga
$p$--orbital in GaAs was ionized as deduced from a partial charge
analysis at the VBM.

In order to find which orbital should be corrected, we used the new
\textsc{pes} module in
\textsc{WIEN2k}.\cite{bagheriDFTCalculationsEnergy2019} Using this
module, we can decompose the interstital charge into their atomic
orbital contributions and get atomic partial charges uniquely and
independently on the choice of the atomic sphere radii and the
localization of different orbitals. For instance in GaN only
$16.2\,\%$ of the Ga-4$p$ charge, but $97.5\,\%$ of Ga-3$d$ charge are
enclosed inside the atomic sphere, and thus the Ga-3$d$ charge
dominates over Ga-4$p$ when considering the charges within the atomic
sphere. However, the rescaled orbital character contributions at the
VBM are $12.1\,\%$ and $8.2\,\%$ of Ga-4$p$ and Ga-3$d$, respectively,
and $79.8\,\%$ of N-2$p$. For the heavier Ga$X$ compounds, we find
progressively larger Ga-$p$ and smaller Ga-$d$ character contributions
at the VBM. Thus, that means that a proper ionization correction for the Ga
compounds should be applied to the Ga-4$p$ orbital.

\begin{figure}[htbp!]
\includegraphics[width=\columnwidth]{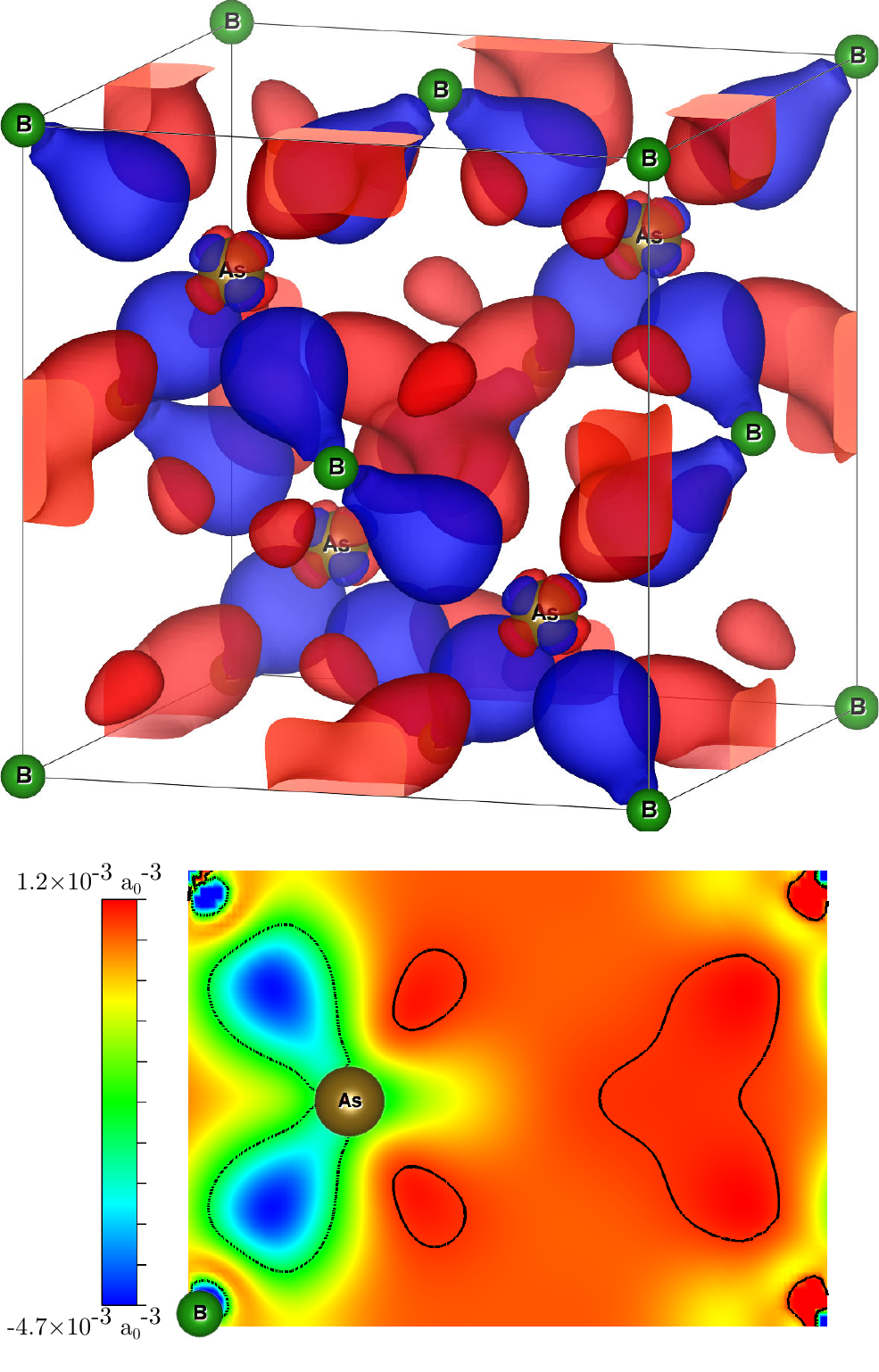}
\caption{Three-dimensional (upper panel) and two-dimensional (lower panel)
plots of the electron density difference $\text{CBM}-\text{VBM}$ in BAs. On
the three-dimensional plot, the positive (CBM, in red) and negative (VBM, in
blue) isosurfaces are defined at \SI{0.9d-3}{\bohr}$^{-3}$ and
\SI{-2d-3}{\bohr}$^{-3}$, respectively. On the two-dimensional plot, the
slice corresponds to a $1\overline{1}0$ plane with B atoms at the corners
and an As atom at the left center, and the black contour lines correspond to
the isosurfaces on the three-dimensional plot.}
  \label{fig:el-dens-BAs}
\end{figure}

The comparison of our calculations to those found in literature needs to be
done carefully, because the correction potential $v_S$ is not always
calculated the same way (e.g., 1/2- or 1/4-ionization and on which atoms)
and, furthermore, the details are not always specified. For instance, our
LDA--1/4 result for Si agrees perfectly with the one from
\textcite{ferreiraApproximationDensityFunctional2008} while in this same
work C was corrected with a 1/4-ionization for both $p$-- and $s$--orbitals,
leading to a value of \SI{5.25}{\electronvolt} that differs substantially
from our result even when we use the same ionization scheme
(\SI{5.87}{\electronvolt}, which is very close to \SI{5.82}{\electronvolt}
with LDA--1/2). This discrepancy for C is unclear.

The comparison with the results from the recent implementation of the
DFT--1/2 method in the LAPW \textsc{exciting}
code\cite{rodriguespelaLDA1MethodImplemented2017} shows perfect agreement,
but it also shows the importance of knowing the exact correction procedure,
since for AlP the agreement is obtained if the correction is applied only to
the P atom, although we cannot be sure that this scheme was used in
Ref.~\onlinecite{rodriguespelaLDA1MethodImplemented2017}. However, for GaN
and GaAs our results in Table~\ref{tab:g4-gaps} (without spin-orbit coupling
for GaAs) show that agreement with those of
\textcite{rodriguespelaLDA1MethodImplemented2017} is only obtained if the Ga
$d$--orbitals (and not the $p$--orbitals) and N/As $p$--orbitals are
corrected, although in GaAs the Ga $p$--orbital contributes non-negligibly
to the VBM. Thus, these examples show that for a meaningful comparison of
results of two sets of DFT--1/2 calculations, one needs to know the details
of the calculations, since depending on the ionization correction (1/4 or
1/2) and on which atoms/orbitals it is applied, a sizeable variation in the
results can be obtained.

In general, a more valid explanation for some of the overestimations found
in covalent materials is that these are not necessarily due to overlapping
holes, but simply due to the fact that the assumptions used in deriving the
method (see Sec.~\ref{sec:theory}) may be too crude. The larger the
difference between the VBM density and the corresponding atomic density
(from which the self-correction potential is calculated) is, the worse the
DFT--1/2 method should perform. This is illustrated with the case of BAs,
where even the 1/4-ionization correction clearly overestimates the
experimental band gap. The VBM in BAs has very little pure atomic character,
but is strongly $sp^{3}$ hybridized and thus very aspherical, as seen on
Fig.~\ref{fig:el-dens-BAs}. The asphericity in the valence distribution
causes an overestimation of the band gap, because the matrix element of
$v_S$ [Eq.~(\ref{eq:8})] will be too large when the charge distribution is
spread out compared to the non-hybridized atomic case. In many cases, this
will also cause an overlap, but not always (see for example BeTe below).

\subsection{Be compounds}
\label{sec:be_coumpounds}

\begin{table*}[htb]
  \caption{Band gaps (in eV) of Be compounds calculated using the DFT--1/2
method with different underlying functionals (LDA or PBE) and ionization
degrees (1/2 or 1/4). In all cases the ionization was applied to the Be
$s$-- and anion $p$--orbitals.  The cutoff radii $r_{c}$ (in $a_{0}$) in
Eq.~(\ref{eq:11}) are indicated in the second column. Only the cutoff radii
from LDA--1/2 calculations are shown, however we checked that those for
PBE--1/2, LDA--1/4, and PBE--1/4 calculations are practically identical (the
difference is below \SI{0.05}{\bohr}, which does not impact the band
gap). For comparison, LDA, PBE, and mBJ results are also shown. The
experimental and $G_{0}W_{0}$ results are from
Refs.~\onlinecite{leePredictionModelBand2016,
yimSynthesisPropertiesBeTe1972, nagelstrasserBandStructureBeTe1998}. The
most accurate values among the DFT--1/2 methods are underlined.}
\begin{ruledtabular}
\begin{tabular}{lcccccccccc}
Solid & $r_{c}$ & LDA & LDA--1/4 & LDA--1/2 & PBE & PBE--1/4 & PBE--1/2 & mBJ & $G_{0}W_{0}$ & Expt. \\
\hline BeO & 0.00, 2.52 & 7.49 & 8.71 & 10.05 & 7.57 & 8.88 & \underline{10.32} & 9.58 & & 10.60 \\
BeS & 0.44, 3.28 & 2.92 & 3.74 & 4.60 & 3.13 & 4.01 & \underline{4.94} & 4.13 & 4.92 & $> 5.5$ \\
BeSe & 0.48, 3.44 & 2.34 & 3.14 & 4.00 & 2.51 & 3.36 & \underline{4.26} & 3.39 & 4.19 & 4.0--4.5\\
BeTe & 0.33, 3.79 & 1.57 & 2.25 & \underline{2.97} & 1.69 & \underline{2.41} & 3.17 & 2.33 & & 2.7 \\
\end{tabular}
\end{ruledtabular}
  \label{tab:BeX}
\end{table*}

An interesting case study for the DFT--1/2 method that has not been
considered previously consists of the Be compounds BeO, BeS, BeSe, and BeTe,
where the first one has the wurtzite structure, while the others have the
zincblende structure. We chose these compounds to investigate the behavior
of DFT--1/2 because of the descending order of ionicity along the
series. The results for the band gap can be found in Table~\ref{tab:BeX}
where we can see that the standard 1/2-ionization correction is the most
effective for the first three compounds. PBE--1/2, for instance, yields a
band gap of \SI{10.32}{\electronvolt} for BeO, which is within a few percent
of the experimental value of \SI{10.60}{\electronvolt}, and band gaps for
BeS (\SI{4.94}{\electronvolt}) and BeSe (\SI{4.26}{\electronvolt}) that
coincide with $G_{0}W_{0}$ results. For BeTe, the PBE--1/2 value is too
large by \SI{0.5}{\electronvolt}, while the errors of
\SI{\sim0.3}{\electronvolt} with PBE--1/4 and LDA--1/2 are somewhat smaller.

\begin{figure*}[htbp!]
\subfloat[]{\includegraphics[width=.99\columnwidth]{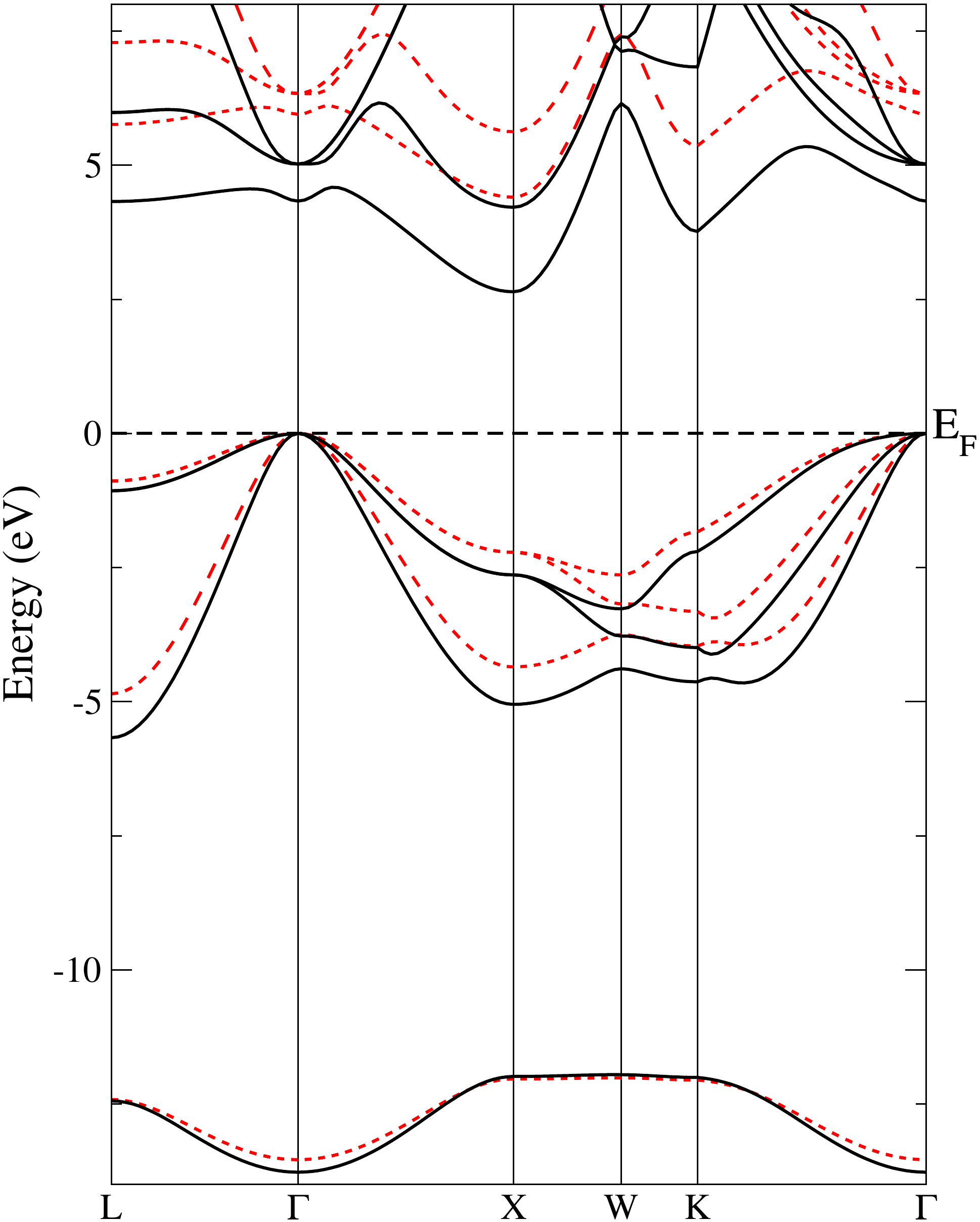}}
  \label{fig:BeSe-band-struc}\hspace{\columnsep}
\subfloat[]{\includegraphics[width=.99\columnwidth]{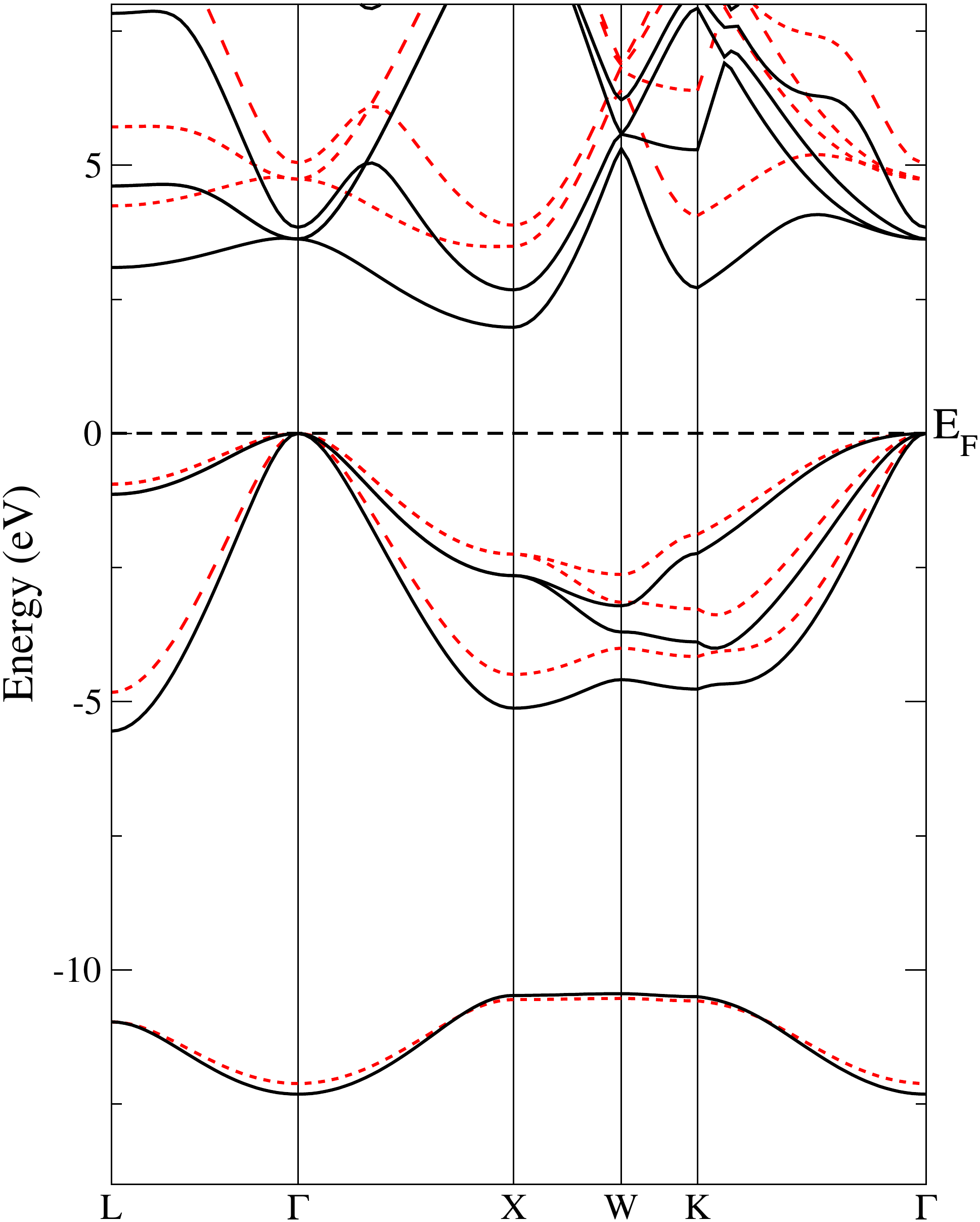}}
  \label{fig:BeTe-band-struc}
\caption{PBE (black solid line) and PBE--1/2 (red dashed line) band
structures for BeSe (a) and BeTe (b). For better visibility, we show band
structures calculated without spin--orbit effects.}
\label{fig:band-struc-Be}
\end{figure*}

In order to investigate the difference between, e.g., BeSe and BeTe, we now
consider the PBE and PBE--1/2 band structures as well as the electron
density close to the Fermi energy. The band structures for both compounds
(see Fig.~\ref{fig:band-struc-Be}) show a very similar change when the
1/2-ionization correction is applied. Compared to PBE, the gap separating
the valence and conduction bands is larger and the bands are more flat. The
shift of the bands is not uniform, but no dramatic change in the shape of
the bands is induced.

\begin{figure*}[htbp!]
\subfloat[]{\includegraphics[width=.99\columnwidth]{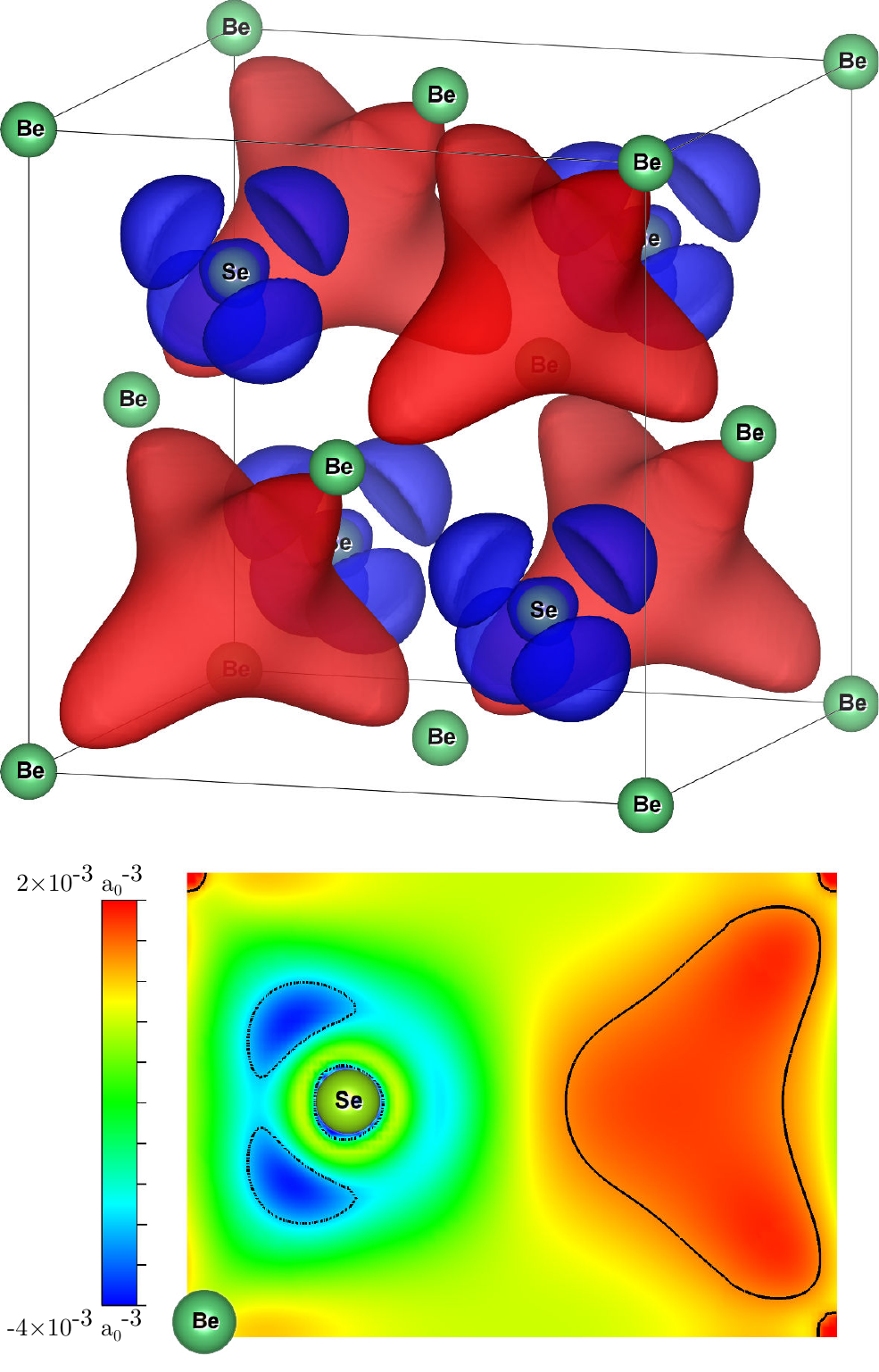}}
  \label{fig:BeSe-diff}\hspace{\columnsep}
\subfloat[]{\includegraphics[width=.99\columnwidth]{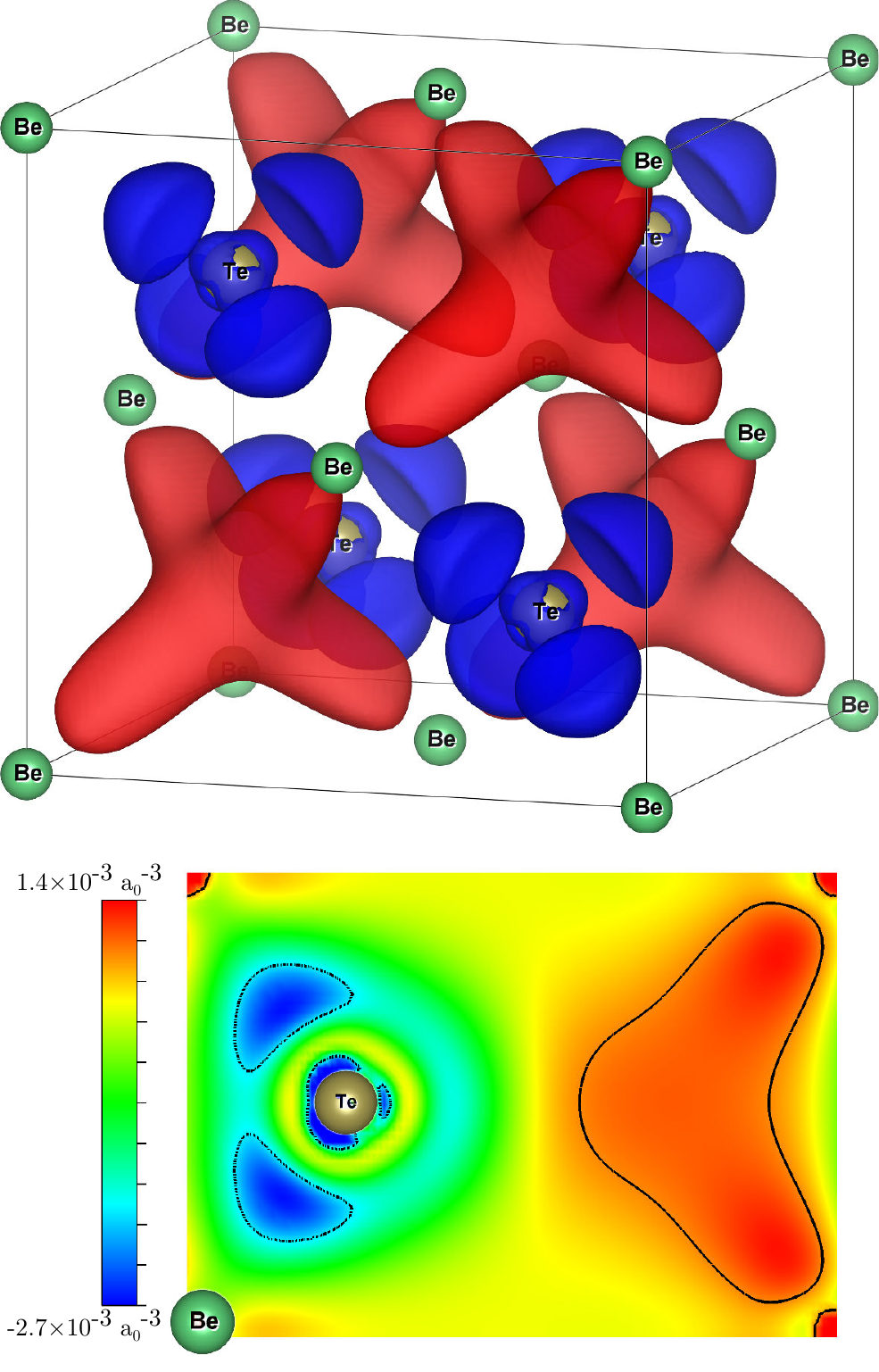}}
  \label{fig:BeTe-diff}
\caption{Three-dimensional (upper panel) and two-dimensional (lower panel) plots of the
electron density difference $\text{CBM}-\text{VBM}$ in BeSe (a) and BeTe
(b). On the three-dimensional plots, the positive (CBM, in red) and one
negative (VBM, in blue) isosurfaces are defined at
\SI{1.25d-3}{\bohr}$^{-3}$ and \SI{-2.75d-3}{\bohr}$^{-3}$ for BeSe and at
\SI{0.9d-3}{\bohr}$^{-3}$ and \SI{-1.75d-3}{\bohr}$^{-3}$ for BeTe. On the
two-dimensional plots, the slice corresponds to a $1\overline{1}0$ plane
with Be atoms at the corners and the anion atom at the left center, and the
black contour lines correspond to the isosurfaces on the three-dimensional
plots.}
\label{fig:el-dens-Be}
\end{figure*}

Figure \ref{fig:el-dens-Be} shows plots of electron density difference
between the VBM and the CBM that are calculated in a small energy
range above the CBM and below the VBM respectively, while ensuring
that the total charge in both cases is equal. Two isosurfaces are
shown; one in red with a positive sign (corresponding to the CBM) and
one in blue (corresponding to the VBM). In the case of the Be
compounds this is almost equivalent to simply superposing both
densities in different colors, because both are well separated
spatially (which is not always true, see
Sec.~\ref{sec:trans-metal-oxid} below). The main observation is that
there are no distinctive features that could be used to clearly judge
a priori which correction (1/2 or 1/4) would be most
suitable. Moreover, in both cases the valence density is mostly
distributed around the anion. This is reflected in the values of the
cutoff radii $r_{c}$ of Be, which in both cases is optimized to very
small values (see Table~\ref{tab:BeX}), such that the correction
potential on the cation is therefore negligible. Thus, in the case of
BeTe, overlapping holes can not explain the overestimation of the band
gap in PBE--1/2. Also, a partial charge analysis (again using the
\textsc{pes} module) of the VBM reveals a nearly identical atomic
$p$--orbital character of the anion in BeSe and BeTe of $96.5\,\%$ and
$94.3\,\%$, respectively.

\subsection{Transition-metal oxides}
\label{sec:trans-metal-oxid}

\begin{table*}[htb]
  \caption{Band gaps (in eV) of TM oxides calculated using the DFT--1/2
    method with different underlying functionals (LDA or PBE). In all cases
    the ionization was applied to the TM $d$ and O $p$. The cutoff radii
    $r_{c}$ (in $a_{0}$) in Eq.~(\ref{eq:11}) are indicated in the second
    column. Only the cutoff radii from LDA--1/2 calculations are shown,
    however we checked that those for PBE--1/2, LDA--1/4, and PBE--1/4
    calculations are practically identical (the difference is below
    \SI{0.05}{\bohr}, which does not impact the band gap). For comparison
    purposes, LDA, PBE, and mBJ results are also shown. The experimental
    results are from Refs.~\onlinecite{leePredictionModelBand2016,
      hummelshojCommunicationsElementaryOxygen2010,
      gillenAccurateScreenedExchange2013,wangElectronicStructuresMathrmC2016}.
    The most accurate values among the DFT--1/2 methods are underlined.}
  \begin{ruledtabular}
    \begin{tabular}{lccccccc}
      Solid & $r_{c}$ & LDA & LDA--1/2 & PBE & PBE--1/2 & mBJ & Expt. \\
\hline
TiO$_{2}$ & 0.29, 2.76 & 1.80 & 3.16 & 1.89 & \underline{3.38} & 2.56 & 3.30 \\
VO$_{2}$ & & metal & metal & metal & metal & 0.51 & 0.6 \\
Cu$_{2}$O & 2.73, 2.21 & 0.53 & 1.09 & 0.53 & \underline{1.14} & 0.82 & 2.17 \\
ZnO & 1.68, 2.80 & 0.74 & 3.26 & 0.81 & \underline{3.50} & 2.65 & 3.44 \\
Cr$_{2}$O$_{3}$ (AFM) & 0.24, 2.0 & 1.20 & 1.35 & 1.64 & \underline{1.76} & 3.68 & 3.4 \\
MnO (AFM) & 1.44, 2.90 & 0.74 & 1.89 & 0.89 & \underline{2.33} & 2.94 & 3.9 \\
FeO (AFM) & & metal & metal & metal & metal & 1.84 & 2.4 \\
Fe$_{2}$O$_{3}$ (AFM) & 0.35, 2.87 & 0.33 & 1.33 & 0.56 & \underline{1.66} & 2.35 & 2.2 \\
CoO (AFM) & 1.72, 2.57 & metal & metal & metal & \underline{0.17} & 3.13 & 2.5 \\
NiO (AFM) & 1.35, 2.17 & 0.43 & 0.66 & 0.95 & \underline{1.33} & 4.14 & 4.3 \\
CuO (AFM) & 5.42, 2.10 & metal & 0.84 & 0.06 & \underline{1.17} & 2.27 & 1.44 \\
\end{tabular}
\end{ruledtabular}
\label{tab:TM}
\end{table*}

Another class of materials that provides a challenge for the DFT--1/2 method
are TM oxides. Results for some representative nonmagnetic and AFM cases are
shown in Table~\ref{tab:TM}. For all TM oxides the correction is based on a
1/2-ionization of the TM atom $d$--orbital and O $p$--orbital. For the AFM
systems the self-energy correction potential $v_S$ is spin-dependent (and
respects the AFM ordering) and calculated from a 1/2-ionization of the spin
with the largest contribution to the VBM.

From the results we can see that the DFT--1/2 band gaps for the nonmagnetic
TiO$_2$ and ZnO are much larger than LDA/PBE and very close to experiment
(errors are below \SI{0.2}{\electronvolt}). However, for all other systems
the band gaps calculated using DFT--1/2 are still much smaller than
experiment. Actually, the DFT--1/2 method works very well for TiO$_2$ and
ZnO since these two systems have a charge-transfer band gap and thus a clear
spatial separation between the VBM and CBM. This is not the case for the
other systems (Cu$_2$O, VO$_2$, and the AFM strongly correlated systems),
where a significant $d$--character is present in both the VBM and CBM, such
that the spherical atomic self-energy correction $v_S$ can not really
distinguish these split bands; any correction that is applied to the VBM
will also influence the energy level of the CBM in a similar way and thus
fails to increase the band gap as much as one would like. It is well known
that the standard (semi)local functionals like PBE are not able to describe
strongly correlated systems properly even at the qualitative
level,\cite{TerakuraPRB84} and only more advanced methods like DFT+$U$,
hybrid functionals, or mBJ lead can lead to reasonable
results. \cite{AnisimovPRB91,TranPRB06,MarsmanJPCM08,TranPRM18,
gerosaAccuracyDielectricdependentHybrid2018} Note, however, that although
mBJ performs better than DFT--1/2 overall, it fails for some cases, notably
TiO$_2$, Cu$_2$O, and ZnO (see Table~\ref{tab:TM} and
Ref.~\onlinecite{kollerMeritsLimitsModified2011}).

In more detail, for MnO, Fe$_2$O$_3$, and CuO, DFT--1/2 leads to a clear
(but not sufficient) improvement of at least \SI{1}{\electronvolt} in the
band gap, which is, however, not as impressive as in TiO$_2$ and ZnO. This
has to be compared to the small improvement of a few tenths of an eV
obtained for NiO and Cr$_2$O$_3$ and the metallic character that persists
for FeO, CoO, and VO$_{2}$. Actually, the common feature of Fe$_2$O$_3$,
MnO, and CuO is to have CBM and VBM that are made of states of opposite
spins, as a consequence of the large exchange splitting.\cite{TerakuraPRB84}
Since the correction potential $v_S$ is spin-dependent (the ionization is
done for the spin with the largest contribution to the VBM), a sizeable
increase in the band gap is possible.  In other TM oxides like NiO, FeO, or
CoO the crystal field splitting is dominant, such that the VBM has a more
mixed spin population.

\begin{figure}[htbp!]
\includegraphics[width=\columnwidth]{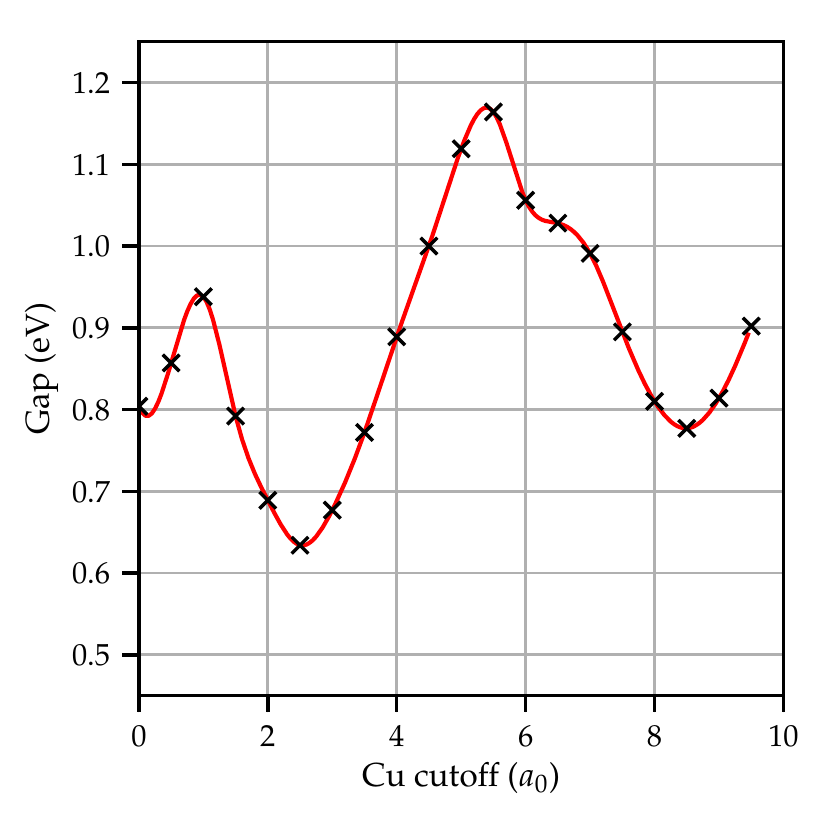}
  \caption{CuO band gap using PBE--1/2 as a function of the cutoff radius of
Cu. Splines (red line) are fitted through the calculated results (black
crosses).}
  \label{fig:CuO-cut-curve}
\end{figure}

In CuO, the choice for the spin for calculating the potential $v_{S}$ is
crucial. The calculation done with the ionization correction applied to the
correct spin (i.e. the one which has the largest population at the VBM for
the given atom) gives a PBE--1/2 result of \SI{1.17}{\electronvolt}, whereas
calculating $v_{S}$ using the other spin results in a band gap of only
\SI{0.49}{\electronvolt}. Another particular feature of CuO is the cutoff
radius $r_{c}$ of the Cu atom, which is extremely large
(\SI{5.42}{\bohr}). As shown in Ref.~\onlinecite{xueImprovedLDA1Method2018},
the band gap as a function of $r_{c}$ may consist of several maxima when
$r_{c}$ reaches the next coordination shell. Usually, one would expect the
global maximum of the band gap with respect to $r_{c}$ to be at the first
maximum (see Ref.~\onlinecite{xueImprovedLDA1Method2018}), however, as shown
in Fig.~\ref{fig:CuO-cut-curve} the second maximum at \SI{5.42}{\bohr} is
higher than the first one at about \SI{1}{\bohr}. Note that $r_{c} =
\SI{5.42}{\bohr}$ is very close to the distance to the nearest Cu atom of
\SI{5.48}{\bohr}, which means an overlap with a large portion of the
neighboring Cu $d$--orbitals. Such overlap introduces a small anisotropy in
the superimposed correction potential $v_S$ around each Cu atom. We mention
that because of numerical problems in the calculations when cutoff radii
larger than \SI{10}{\bohr} are used, we could not verify whether the third
local maximum would be even higher or not.

Interestingly, in Fe$_2$O$_3$ the reverse is observed. A slightly larger
band gap (\SI{1.87}{\electronvolt} with PBE--1/2) is obtained when the wrong
spin is ionized for calculating the correction potential. We hypothesize
that this behavior is caused by a larger bonding-antibonding splitting of
the Fe-$d_{e_{g}}$$-$O-$p$ interaction when the wrong spin is chosen.

\begin{figure*}[htbp!]
\subfloat[]{\includegraphics[width=.99\columnwidth]{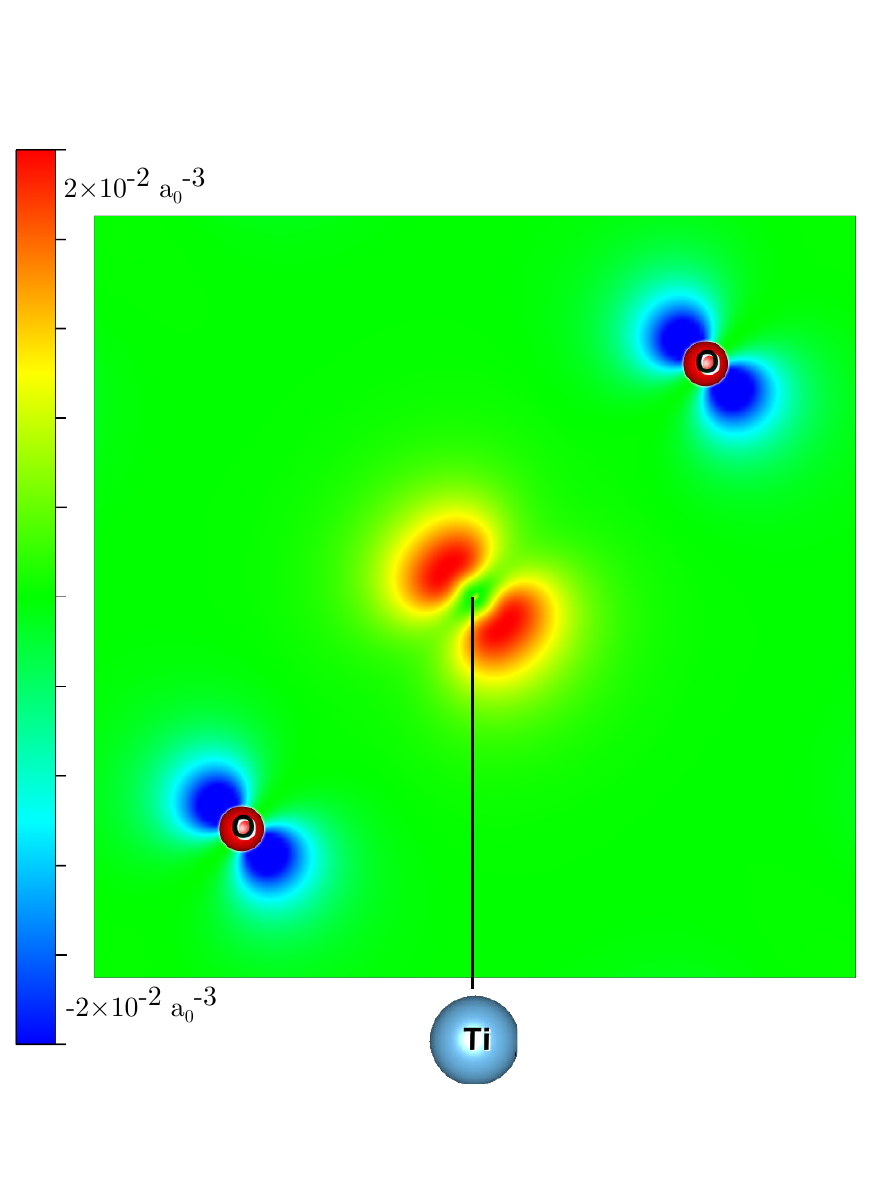}}
  \label{fig:tio2-diff-dens}\hspace{\columnsep}
\subfloat[]{\includegraphics[width=.99\columnwidth]{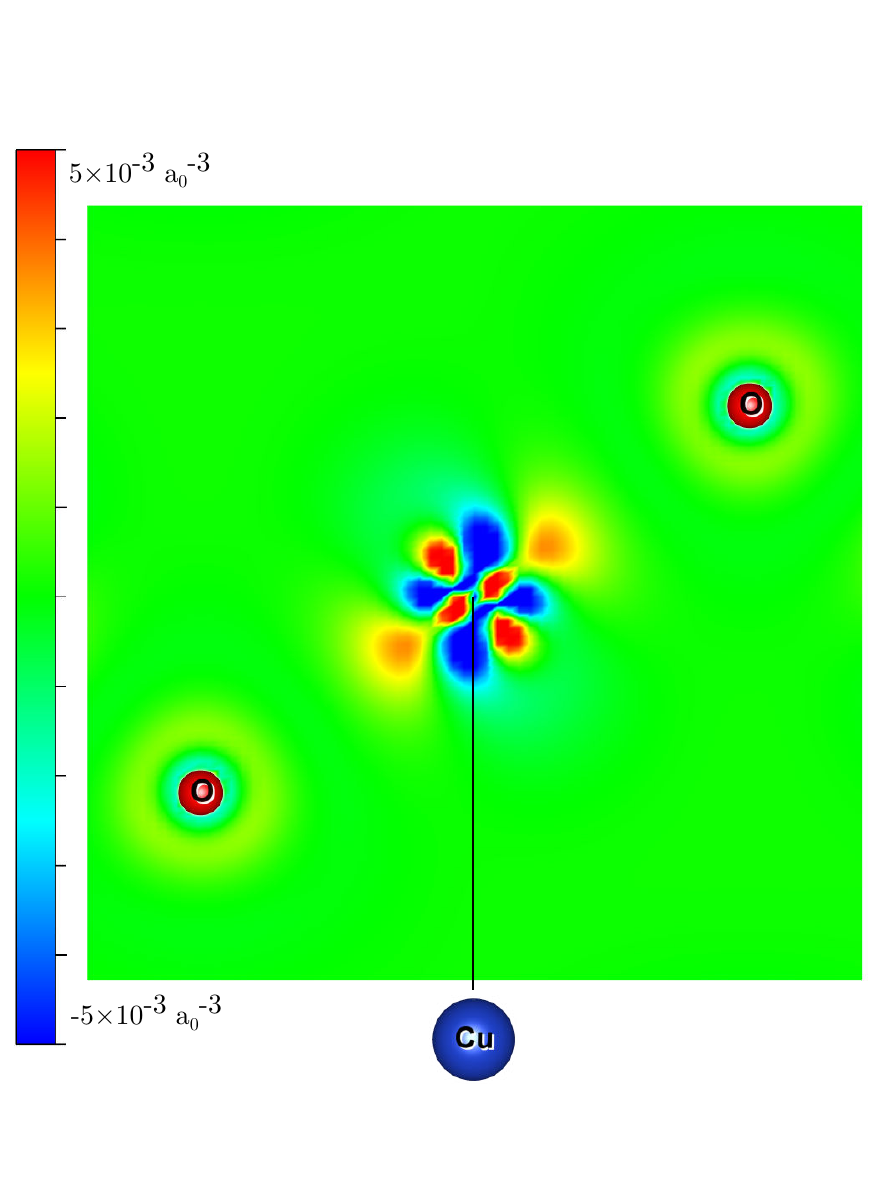}}
  \label{fig:cu2o-diff-dens}
  \caption{Density difference ($\text{CBM}-\text{VBM}$) plots for TiO$_2$
(a) and Cu$_2$O (b). For TiO$_2$ the $001$ plane is shown, with a Ti atom at
the center and two nearest-neighbor O atoms at the corners. For Cu$_2$O, the
$1\overline{1}0$ plane is shown, with a Cu atom at the center and two
nearest-neighbor O atoms at the corners.}
  \label{fig:diff-dens-tmo}
\end{figure*}

Now, a comparison between TiO$_2$ (accurately described by DFT--1/2) and
Cu$_2$O (inaccurately described by DFT--1/2) is
made. Figure~\ref{fig:diff-dens-tmo} shows difference density plots, where
the density around the VBM is subtracted from the one around the CBM. In
TiO$_2$ the VBM and CBM are, as expected, spatially well separated with the
conduction band consisting primarily of the Ti $d$--orbitals and the valence
band of the O $p$--orbitals. In Cu$_2$O, however, both bands are
predominantly composed of Cu $d$--orbitals, i.e., the $d$--orbitals are
split across the Fermi level due to the crystal field. The conduction band
(in red) has a strong $d_{z^2}$ character with lobes pointing towards the O
atom, while the lobes of the valence band (in blue) point in other
directions. Thus, as clearly visible, in Cu$_{2}$O the VBM and CBM are
located on the same atom such that the spherical correction potential $v_S$
can barely increase the energy difference between the VBM and CBM.

\begin{figure*}[htbp!]
\subfloat[]{\includegraphics[width=.99\columnwidth]{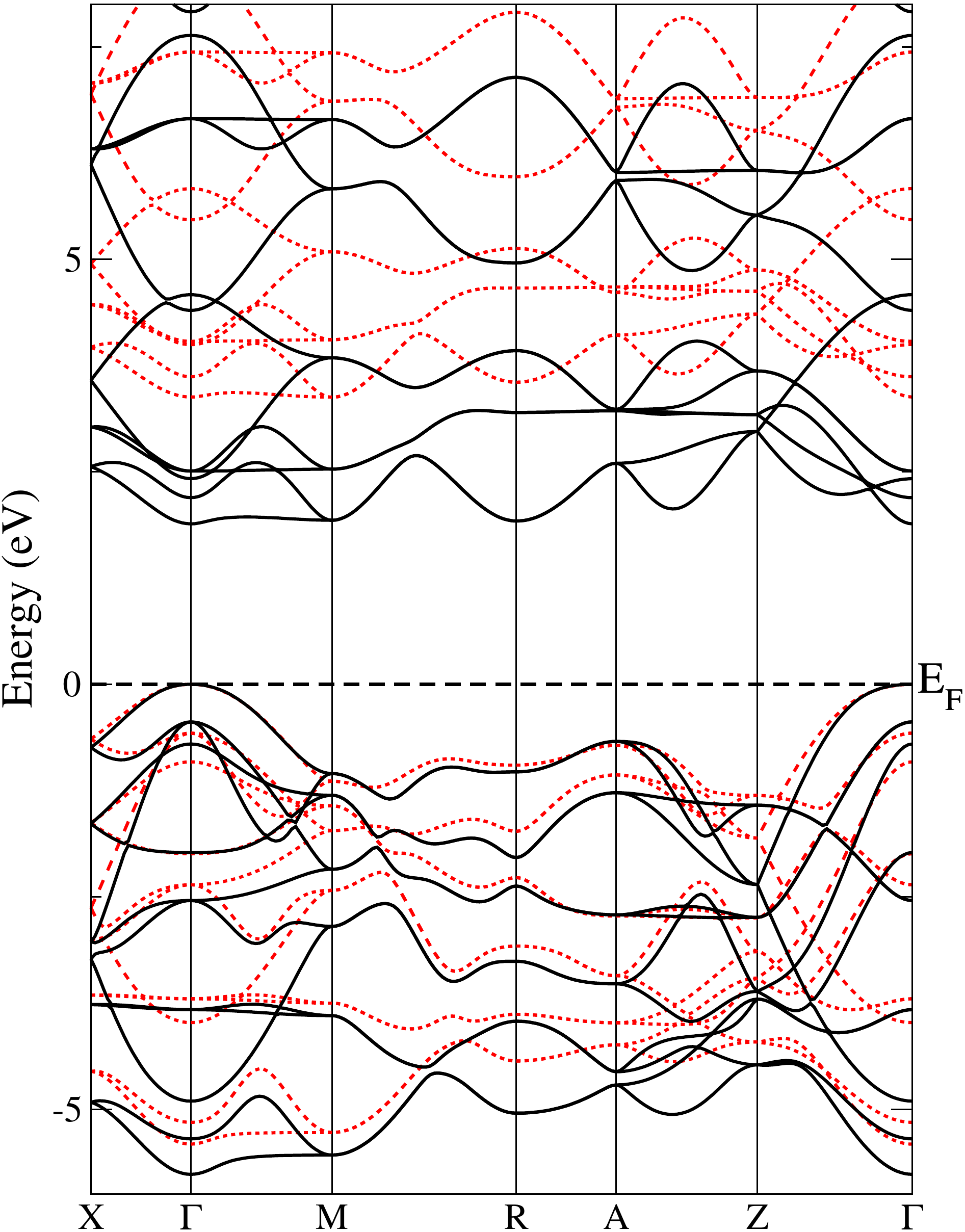}}
  \label{fig:tio2-band-struc}\hspace{\columnsep}
\subfloat[]{\includegraphics[width=.99\columnwidth]{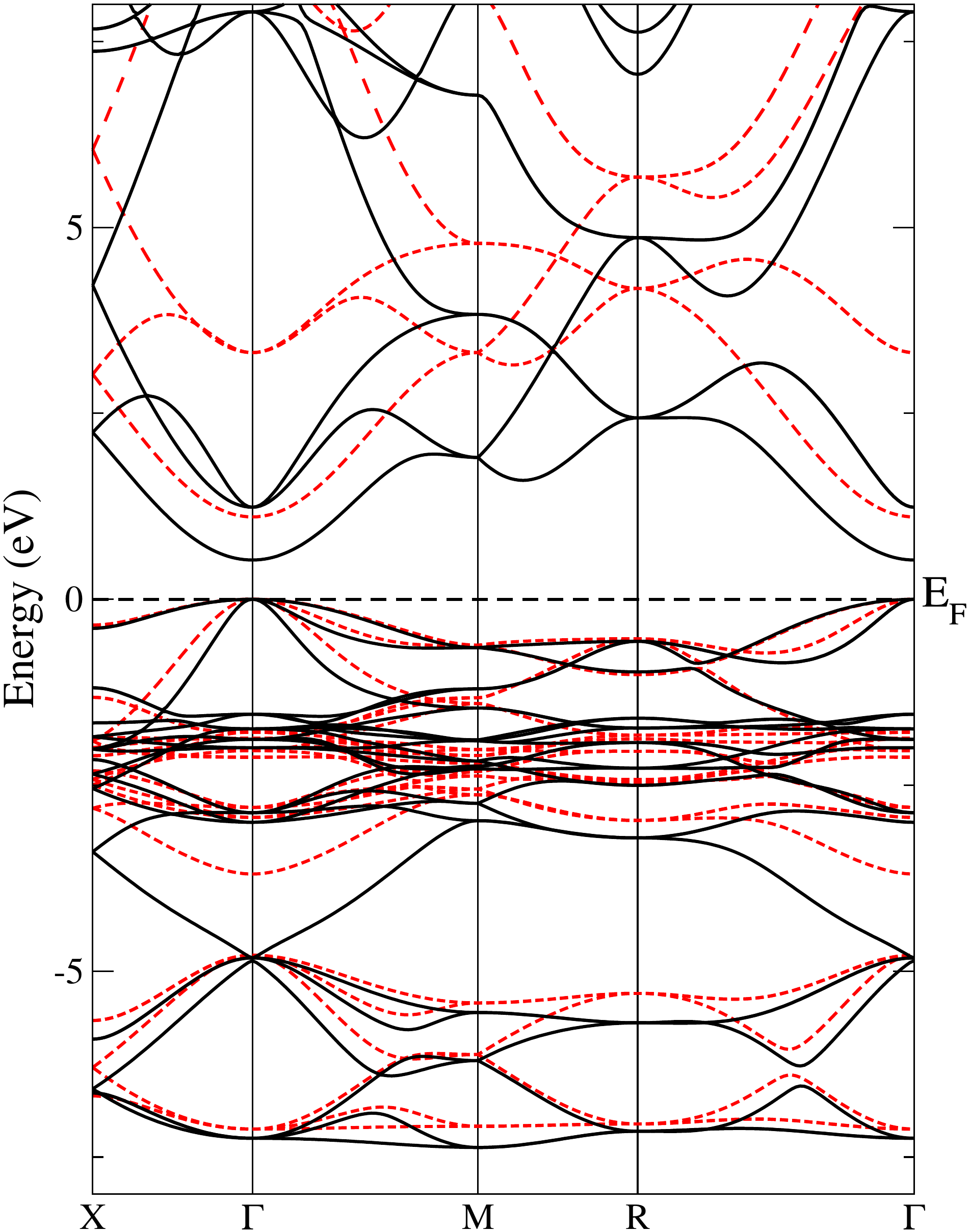}}
  \label{fig:cu2o-band-struc}
\caption{PBE (black solid line) and PBE--1/2 (red dashed line) band
structures for TiO$_{2}$ (a) and Cu$_{2}$O (b).}
\label{fig:band-struc-tmo}
\end{figure*}

The PBE and PBE--1/2 band structures of TiO$_2$ and Cu$_{2}$O are shown in
Figs.~\ref{fig:band-struc-tmo}(a) and \ref{fig:band-struc-tmo}(b),
respectively. For TiO$_2$, the band gap is approximately twice as
large. Changes in the shape of the bands are rather minor for the conduction
bands, but more pronounced differences can be observed in the occupied
bands, e.g., at the $\Gamma$ and R points in the range 4 --
\SI{5}{\electronvolt} below the Fermi energy, where the changes do not
consist of a simple shift. Among the differences in the shape of the bands
in Cu$_2$O, there is for instance the crossing of bands at $\Gamma$ at
\SI{-5}{\electronvolt} with PBE, while they are clearly separated with
PBE--1/2. The band that is significantly raised in energy has a strong Cu
$d$--character, whereas the bands that are not shifted relative to the Fermi
energy have strong O $p$--character.

\begin{figure}[htbp!]
\includegraphics[width=\columnwidth]{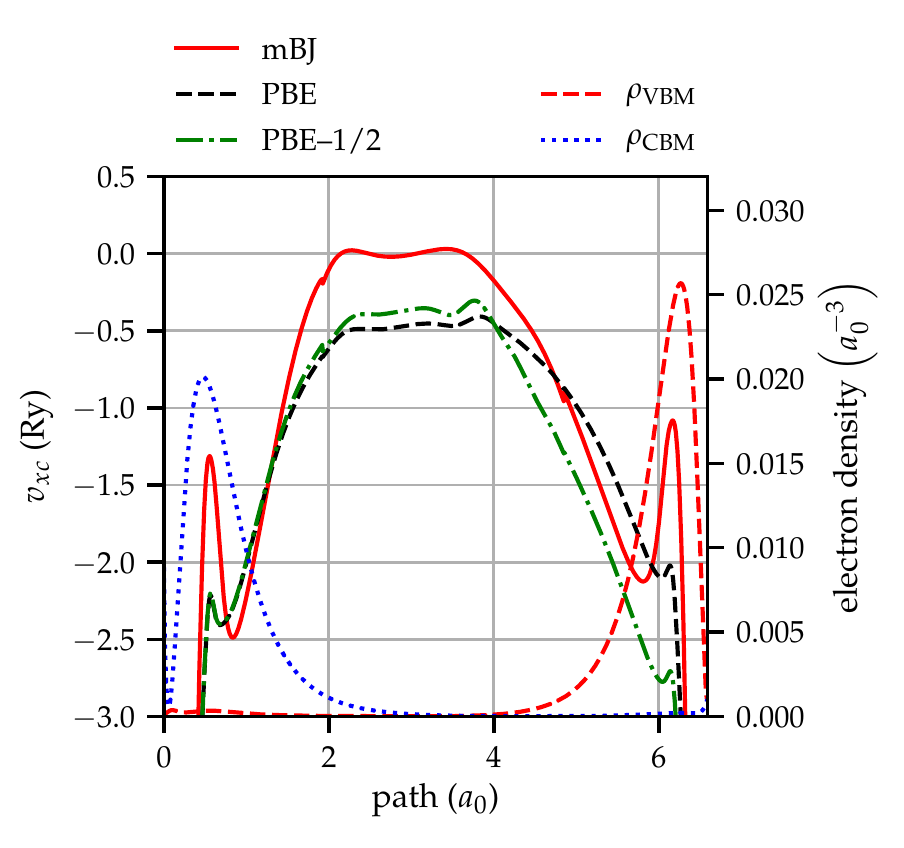}
  \caption{Plots of mBJ, PBE, and PBE--1/2 exchange--correlation potentials
$v_{xc}$ in TiO$_{2}$ and densities of the VBM and CBM. The path is from (0,
1, 0) to (0.305, 0.305, 0) in the unit cell fractional coordinates, thus
from the Ti atom through the interstitial region and terminating at an O
atom. The densities are taken from the PBE calculation.}
  \label{fig:TiO2-vxc}
\end{figure}

Figure~\ref{fig:TiO2-vxc} shows the mBJ, PBE, and PBE--1/2
exchange--correlation potentials in TiO$_{2}$, which lead to band gaps of
\SI{2.56}{\electronvolt}, \SI{1.89}{\electronvolt}, and
\SI{3.38}{\electronvolt}, respectively. As mentioned above, mBJ performs
badly. We can see that compared to PBE, mBJ raises the energy in the
interstitial and has peaks at the outer atomic orbitals for both Ti and O
(where respectively the CBM and VBM have a large density). With PBE--1/2 a
much more accurate band gap is achieved thanks to a significantly more
negative potential at the VBM region around the O atom.

To finish, we mention that \textcite{xueImprovedLDA1Method2018} reported a
similar issue in Li$_2$O$_2$ as in Cu$_{2}$O. In this case the O $p$--bands
are split across the Fermi level, with the VBM formed by the (degenerate)
$p_x$ and $p_y$--orbitals and the CBM by the $p_z$--band, while the
correction potential $v_{S}$ is calculated from an atomic calculation, which
is spherically symmetric. Thus, as in
Ref.~\onlinecite{xueImprovedLDA1Method2018}, a severe underestimation of the
band gap for Li$_2$O$_2$ is obtained and our LDA--1/2 and PBE--1/2 values
are 2.52 and \SI{2.71}{\electronvolt}, respectively (only
\SI{\sim0.5}{\electronvolt} larger than LDA and PBE), while experiment is
\SI{4.91}{\electronvolt}. With a value of \SI{4.81}{\electronvolt}, the mBJ
potential succeeds in describing the band gap very accurately.

\subsection{Shell correction for DFT--1/2}
\label{sec:shell-correction-lda}

\begin{figure}[htbp!]
\subfloat[]{\includegraphics[width=\columnwidth]{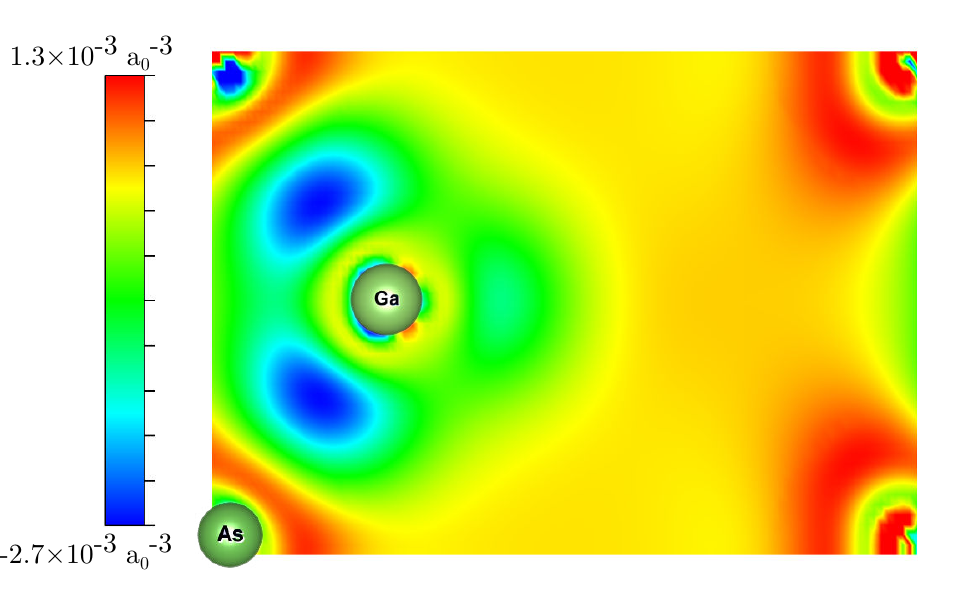}}\\
\subfloat[]{\includegraphics[width=0.9513\columnwidth]{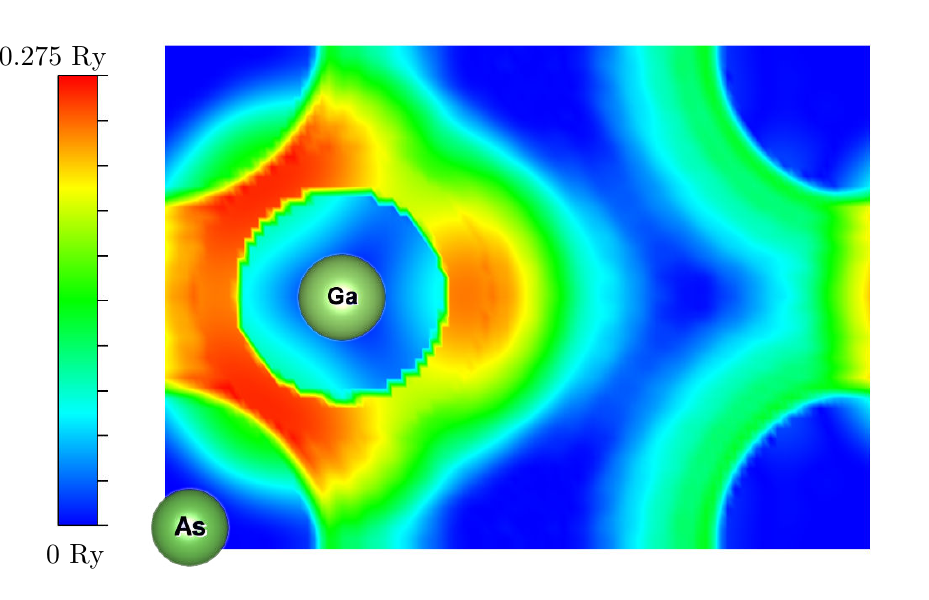}}
  \caption{Density difference ($\text{CBM}-\text{VBM}$) (a) and shPBE--1/4
potential $v_S$ (b) for GaAs. The correction potentials of both atoms
spatially overlap with the valence density maximum around the As atom. The
procedure optimizes the four radial parameters ($r_{\text{in}}$ and
$r_{\text{out}}$ of both atoms) such that the gap is maximized. This clearly
means that the overlap of the total correction potential with the VBM
density will be large while it will be minimized with the CBM density.}
  \label{fig:gaas-shell-vs}
\end{figure}

\textcite{xueImprovedLDA1Method2018} proposed a more general version of
DFT--1/2, called shDFT--1/2 (sh is a shorthand for shell), which employs a
modified, shell-like cutoff function
\begin{align} \Theta \left( r \right) & =
                           \begin{cases} \left( 1 - \left[ \frac{2 \left( r
-r_{\mathrm{in}} \right)}{r_{\mathrm{out}} - r_{\mathrm{in}}} -1
\right]^{20} \right)^3 \qquad & \text{$r_{\mathrm{in}} < r <
r_{\mathrm{out}}$} \label{eq:12} \\ 0 \qquad & \text{otherwise}
                         \end{cases}
\end{align} with two variationally determined parameters $r_{\mathrm{in}}$
and $r_{\mathrm{out}}$ and a sharper cutoff compared to
Eq.~(\ref{eq:11}). Note that the shell-like cutoff function reduces to the
spherical one of Eq.~(\ref{eq:11}) when the inner radius is chosen as
$r_{\mathrm{in}} = - r_{\mathrm{out}}$, but with an exponent of 20 instead
of 8. The inner radius $r_{\mathrm{in}}$ is also used to maximize the band
gap, which implies that the band gap calculated with shDFT--1/2 should be
larger compared to optimizing only $r_{\text{out}}$ as done with
DFT--1/2. The aim of introducing an inner radius is to avoid unwanted
interaction of (semi-)core electrons with the correction potential
$v_{S}$. However, optimizing the radii in the shDFT--1/2 method is more
tedious, since $r_{\mathrm{in}}$ and $r_{\mathrm{out}}$ need to be optimized
simultaneously and may be interdependent. For example in GaAs, DFT--1/2
requires a Ga cutoff of
$\SI{1.23}{\bohr}$,\cite{ferreiraApproximationDensityFunctional2008} whereas
shLDA--1/4 requires $r_{\mathrm{in}} = \SI{2.1}{\bohr}$, and
$r_{\mathrm{out}} = \SI{3.9}{\bohr}$\cite{xueImprovedLDA1Method2018} for the
same atom. In this case (as in some others) the correction potential of both
atoms overlap with the valence density that is distributed around one of the
atoms as illustrated in Fig.~\ref{fig:gaas-shell-vs} for GaAs. However, a
well-founded explanation why this approach should yield more accurate band
gaps is not provided in Ref.~\onlinecite{xueImprovedLDA1Method2018}.

\begin{table}
\caption{Band gap (in eV) calculated with different cutoff functions:
spherical [Eq.~(\ref{eq:11})] and sharp spherical [Eq.~(\ref{eq:12}) with
$r_{\mathrm{in}}=-r_{\mathrm{out}}$ and $r_{\mathrm{out}}$ determined
variationally]. The reason for choosing this value for $r_{\mathrm{in}}$ is
that the shell-like cutoff function becomes very similar to the spherical
one [Eq.~(\ref{eq:11})], with the only difference being the exponent of the
reduced radius term ${r}/{r_{\mathrm{out}}}$ (see discussion in text). }
\begin{ruledtabular}
\begin{tabular}{lccc} Solid & Eq.~(\ref{eq:11}) & Eq.~(\ref{eq:12}) \\
\hline Ge--1/4 & 0.27 & 0.41 \\ AlP--0--1/2 & 3.21 & 3.33\\ BN--0--1/2 &
6.79 & 6.97 \\ BeSe--0--1/2 & 4.26 & 4.40 \\ GaAs--1/4 & 0.97 & 1.06 \\
\end{tabular}
\end{ruledtabular}
  \label{tab:stepfuncs}
\end{table}

\begin{table*}
  \caption{Band gaps (in eV) of various compounds using the shPBE
method. The cutoff radii $r_{\text{in}}$ and $r_{\text{out}}$ are given for
the specific correction that is required for the compound according to the
rules of \textcite{xueImprovedLDA1Method2018} That means that when two
cutoff radii are given, a 1/4-ionization correction is applied to both
atoms, and when one is given a 1/2-ionization correction is applied only to
the anion. For each of the shPBE calculations the cutoff radii were
optimized for the specific correction, as the cutoff radii between
shPBE--1/4--1/4 and PBE--0--1/2 are not always transferable. For NiO, a
normal 1/2-ionization correction is applied to all atoms in the unit
cell. The most accurate values among the DFT--1/2 methods are underlined.}
\begin{ruledtabular}
  \begin{tabular}{lcccccccc} Solid & $r_{\text{in}}$ & $r_{\text{out}}$ &
   PBE & PBE--1/4 & PBE--0--1/2 & shPBE--1/4 & shPBE--0--1/2 & Expt. \\
    \hline Ge & 1.71 & 3.30 & metal & 0.27 & \underline{0.59} &
    \underline{0.93}\footnotemark[1] & 1.78 & 0.74 \\ SiC & 0.12 & 2.61 &
    1.35 & \underline{2.43} & 3.31 & 2.55 & 3.52\footnotemark[1] & 2.42 \\
    BN & 0.15 & 2.14 & 4.47 & 5.79 & \underline{6.79} & 5.87 &
    7.03\footnotemark[1] & 6.36 \\ BAs & 0.18, 0.89 & 2.81, 2.70 & 1.09 &
    \underline{1.93} & 2.00 & 2.03 & 2.14 & 1.46 \\ AlN & 0.13 & 2.51 & 3.34
    & 4.66 & 5.96 & \underline{4.78} & 6.19 & 4.90 \\ AlP & 0.66 & 3.15 &
    1.59 & \underline{2.50} & 3.21 & 2.59 & 3.36\footnotemark[1] & 2.5 \\
    GaN & 0.16 & 2.53 & 1.66 & 2.55 & \underline{3.41} & 2.71 &
    3.55\footnotemark[1] & 3.28 \\ GaAs & 2.17, 1.48 & 4.21, 3.25 & 0.43 &
    0.97 & \underline{1.49} & \underline{1.54}\footnotemark[1] & 2.04 & 1.52
    \\ BeSe & 0.83 & 2.89 & 2.51 & 3.36 & \underline{4.26} & 3.55 &
    \underline{4.44}\footnotemark[1] & 4.0-4.5 \\ BeTe & 1.15 & 3.22 & 1.69
    & 2.41 & 3.17 & \underline{2.69} & 3.42\footnotemark[1] & 2.7 \\ ZnO &
    0.13, 0.10 & 1.43, 2.38 & 0.81 & 2.08 & \underline{2.32} & 2.21 &
    \underline{2.36}\footnotemark[1] & 3.44\\ NiO & -0.25, 0.12 & 1.29, 1.25
    & 0.95 & metal & \underline{1.33} & metal &
     \underline{1.36}\footnotemark[1] & 4.3 \\
\end{tabular}
\end{ruledtabular} \footnotetext[1]{Obtained with the preferred correction
according to \textcite{xueImprovedLDA1Method2018}}
  \label{tab:shell}
\end{table*}

\textcite{xueImprovedLDA1Method2018} also prescribed a procedure to choose
the correction. In monoatomic compounds, there is a choice to apply either a
1/2- or a 1/4-ionization correction. The former should be used when the VBM
density is distributed around the atom (like in diamond), while the latter
should be used when the VBM density is distributed around the bond center
(like in Si, see Sec.~\ref{sec:group-iv-se}). In binary compounds, either a
1/2-ionization correction is applied to the anion or a 1/4 ionization to
both the anion and cation. Which one of these two corrections is applied
depends on the CBM density distribution. When the CBM density is distributed
close to the cation-cation bonds (AlP is the example given by
\textcite{xueImprovedLDA1Method2018}, but this would apply also to BeSe and
BeTe, see Fig.~\ref{fig:el-dens-Be}), only the anion should be corrected by
a 1/2-ionization. However, when the CBM density is distributed around the
atoms, like in GaAs (Fig.~\ref{fig:gaas-shell-vs}), a 1/4-ionization
correction should be applied to both atoms, and in such a case a large
$r_{\mathrm{in}}$ should minimize the interaction of the correction
potential $v_S$ with the CBM. However, how to deal with a case like ZnO
where a 1/2-ionization correction on both atoms is needed to obtain a
reasonable band gap \cite{ferreiraSlaterHalfoccupationTechnique2011} is not
discussed. It is also clear that other situations exist, like BAs (see
Fig.~\ref{fig:el-dens-BAs}) where the CBM density is distributed along the
cation bonds but also around the anion (note the CBM lobes around the As
atom, which are absent in BeSe and BeTe, see Fig.~\ref{fig:el-dens-Be}).

Before discussing the results obtained with shPBE--1/2, the influence of the
steepness of the outer part of Eq.~(\ref{eq:12}) is now discussed. As noted
above, the outer cutoff is sharper in the shell function [Eq.~(\ref{eq:12})]
than in the original spherical function [Eq.~(\ref{eq:11})]. In order to
test the influence of the outer steepness on the results, calculations with
Eq.~(\ref{eq:12}) were done using no inner cutoff (i.e., with
$r_{\mathrm{in}}=-r_{\mathrm{out}}$, see discussion above) and the results
are compared to those obtained with Eq.~(\ref{eq:11}). The band gaps
obtained with the two cutoff functions are shown in
Table~\ref{tab:stepfuncs}, where we can see that Eq.~(\ref{eq:12}) leads to
values that are moderately larger by \SIrange{0.1}{0.2}{\electronvolt}. This
is easily explained by noting that a steeper cutoff can more effectively
maximize (minimize) the overlap of $v_S$ with the VBM (CBM).

Representative compounds were considered for calculations with the
shDFT--1/2 method. We chose border cases (BeTe and BeSe), some of the group
IV and III-V semiconductors whose band gaps are significantly underestimated
in a 1/4-ionization correction, and BAs to check if the overestimation found
even in LDA--1/4 is worsened or not. We also included one nonmagnetic (ZnO)
and one AFM TM oxide (NiO) to see the influence of the inner cutoff on this
class of materials. We limited our calculations to the PBE functional.

The results obtained with the shPBE--1/2 methods are shown in
Table~\ref{tab:shell}. Compared to the corresponding PBE--1/2 methods with
the same ionization correction, the improvement is in most cases rather
small or non-existent. Actually, comparing the results to those from
Table~\ref{tab:stepfuncs} discussed above, in most cases (e.g., AlP, BN, or
BeSe) the increase in the band gap is mostly due to the sharper cutoff and
not to the inner radius $r_{\text{in}}$. It is only for Ge and GaAs that the
inner radius has a large influence on the results. For these two latter
cases, excellent agreement with experiment is obtained with shPBE--1/4. The
other main observation is that shPBE-0-1/2 (1/2-ionization correction
applied only to the anion) strongly overestimates the band gap in all cases
except ZnO and NiO. In the case of NiO, it is expected since also the shell
correction can not capture the $d$--$d$ transition that makes up the
fundamental gap. The case of ZnO shows that sometimes a 1/2-ionization
correction on both atoms is required to obtain a good band gap (see
Table~\ref{tab:TM}).

Actually, with the larger set of solids used by
\textcite{xueImprovedLDA1Method2018} to test shLDA--1/2, the overall
improvement is rather modest, in particular when taking account the fact
that an extra parameter ($r_{\text{in}}$) is introduced and leads to a more
cumbersome procedure.

\section{Summary}
\label{sec:summary}

Since the DFT--1/2 method has been proposed, a large number of works
reporting accurate results for the band gap have been published. However, as
discussed in \textcite{xueImprovedLDA1Method2018} and in the present work,
the method has flaws which prevent its straightforward application.
Firstly, for the cases where the states around the band gap, i.e., both at
the VBM and CBM, come from orbitals centered at the same atom, the method
will most likely fail. Such examples discussed in this work are many TM
oxides, but also Li$_2$O$_2$.

Secondly, the method can not be blindly applied to covalent semiconductors
and it is only recently\cite{xueImprovedLDA1Method2018} that this discussion
has been extended beyond the group IV semiconductors. It is rather clear
that there is no \emph{unique} way (1/2- or 1/4-ionization correction, which
atoms, and which orbital) to calculate reliably the band gap for these
materials using (sh)DFT--1/2, without prior knowledge of the experimental
band gap.

The comparison with the mBJ potential shows that mBJ is superior to DFT--1/2
on average. The most visible differences in the performance of DFT--1/2 and
mBJ are for the TM oxides. While DFT--1/2 is very accurate for TiO$_{2}$ and
ZnO, but very inaccurate for the AFM oxides, the reverse is observed with
mBJ.

We also considered the shell correction (shDFT--1/2). It requires the
introduction of an extra parameter, which leads to a more tedious
application of the method. Furthermore, it is only for a few cases that
shDFT--1/2 clearly improves the results.

Thus, we conclude that while DFT--1/2 is a computationally fast method and
can be accurate for band gap calculations, one should be careful in its
application. In particular, the method can be applied efficiently only when
the VBM and CBM are spatially well separated, like in ionic solids, such
that predominantly the VBM is shifted down by the correction potential, and
not the CBM. When these conditions are met, DFT--1/2 is certainly useful
especially in systems with large unit cells, like for example for the
calculations of defect levels,\cite{lucattoGeneralProcedureCalculation2017,
matusalemCombinedLDALDA12013}
surfaces,\cite{belabbesElectronicPropertiesPolar2011} or
interfaces.\cite{ribeiroFirstprinciplesCalculationAlAs2011,
ribeiroAccuratePredictionSi2009} An interesting perspective opened by the
DFT--1/2 technique is the semi-empirical application to larger
structures. One can fit or tune the correction to a reference (e.g. bulk)
configuration by parametrizing either the ionization level, or the
correction factor (multiplying the correction potential by a constant
factor)\cite{ataideFastAccurateApproximate2017,
ribeiroInitioQuasiparticleApproximation2015} and consequently applying this
semi-empirical correction in the structure of interest, e.g. defects,
interfaces or surfaces.

\begin{acknowledgments}

This work was supported by projects F41 (SFB ViCoM), W1243 (Solids4Fun), and
P27738-N28 of the Austrian Science Fund (FWF).

\end{acknowledgments}

\bibliography{LDA_half,DFT_general,TMOs,references}

\begin{thebibliography}{76}%
\makeatletter
\providecommand \@ifxundefined [1]{%
 \@ifx{#1\undefined}
}%
\providecommand \@ifnum [1]{%
 \ifnum #1\expandafter \@firstoftwo
 \else \expandafter \@secondoftwo
 \fi
}%
\providecommand \@ifx [1]{%
 \ifx #1\expandafter \@firstoftwo
 \else \expandafter \@secondoftwo
 \fi
}%
\providecommand \natexlab [1]{#1}%
\providecommand \enquote  [1]{``#1''}%
\providecommand \bibnamefont  [1]{#1}%
\providecommand \bibfnamefont [1]{#1}%
\providecommand \citenamefont [1]{#1}%
\providecommand \href@noop [0]{\@secondoftwo}%
\providecommand \href [0]{\begingroup \@sanitize@url \@href}%
\providecommand \@href[1]{\@@startlink{#1}\@@href}%
\providecommand \@@href[1]{\endgroup#1\@@endlink}%
\providecommand \@sanitize@url [0]{\catcode `\\12\catcode `\$12\catcode
  `\&12\catcode `\#12\catcode `\^12\catcode `\_12\catcode `\%12\relax}%
\providecommand \@@startlink[1]{}%
\providecommand \@@endlink[0]{}%
\providecommand \url  [0]{\begingroup\@sanitize@url \@url }%
\providecommand \@url [1]{\endgroup\@href {#1}{\urlprefix }}%
\providecommand \urlprefix  [0]{URL }%
\providecommand \Eprint [0]{\href }%
\providecommand \doibase [0]{http://dx.doi.org/}%
\providecommand \selectlanguage [0]{\@gobble}%
\providecommand \bibinfo  [0]{\@secondoftwo}%
\providecommand \bibfield  [0]{\@secondoftwo}%
\providecommand \translation [1]{[#1]}%
\providecommand \BibitemOpen [0]{}%
\providecommand \bibitemStop [0]{}%
\providecommand \bibitemNoStop [0]{.\EOS\space}%
\providecommand \EOS [0]{\spacefactor3000\relax}%
\providecommand \BibitemShut  [1]{\csname bibitem#1\endcsname}%
\let\auto@bib@innerbib\@empty
\bibitem [{\citenamefont {Hohenberg}\ and\ \citenamefont
  {Kohn}(1964)}]{hohenbergInhomogeneousElectronGas1964a}%
  \BibitemOpen
  \bibfield  {author} {\bibinfo {author} {\bibfnamefont {P.}~\bibnamefont
  {Hohenberg}}\ and\ \bibinfo {author} {\bibfnamefont {W.}~\bibnamefont
  {Kohn}},\ }\href {\doibase 10.1103/PhysRev.136.B864} {\bibfield  {journal}
  {\bibinfo  {journal} {Phys. Rev.}\ }\textbf {\bibinfo {volume} {136}},\
  \bibinfo {pages} {B864} (\bibinfo {year} {1964})}\BibitemShut {NoStop}%
\bibitem [{\citenamefont {Kohn}\ and\ \citenamefont
  {Sham}(1965)}]{kohnSelfConsistentEquationsIncluding1965}%
  \BibitemOpen
  \bibfield  {author} {\bibinfo {author} {\bibfnamefont {W.}~\bibnamefont
  {Kohn}}\ and\ \bibinfo {author} {\bibfnamefont {L.~J.}\ \bibnamefont
  {Sham}},\ }\href {\doibase 10.1103/PhysRev.140.A1133} {\bibfield  {journal}
  {\bibinfo  {journal} {Physical Review}\ }\textbf {\bibinfo {volume} {140}},\
  \bibinfo {pages} {A1133} (\bibinfo {year} {1965})}\BibitemShut {NoStop}%
\bibitem [{\citenamefont {Perdew}(2009)}]{perdewDensityFunctionalTheory2009}%
  \BibitemOpen
  \bibfield  {author} {\bibinfo {author} {\bibfnamefont {J.~P.}\ \bibnamefont
  {Perdew}},\ }\href {\doibase 10.1002/qua.560280846} {\bibfield  {journal}
  {\bibinfo  {journal} {International Journal of Quantum Chemistry}\ }\textbf
  {\bibinfo {volume} {28}},\ \bibinfo {pages} {497} (\bibinfo {year}
  {2009})}\BibitemShut {NoStop}%
\bibitem [{\citenamefont {Perdew}\ \emph {et~al.}(1996)\citenamefont {Perdew},
  \citenamefont {Burke},\ and\ \citenamefont {Ernzerhof}}]{PerdewPRL96}%
  \BibitemOpen
  \bibfield  {author} {\bibinfo {author} {\bibfnamefont {J.~P.}\ \bibnamefont
  {Perdew}}, \bibinfo {author} {\bibfnamefont {K.}~\bibnamefont {Burke}}, \
  and\ \bibinfo {author} {\bibfnamefont {M.}~\bibnamefont {Ernzerhof}},\
  }\href@noop {} {\bibfield  {journal} {\bibinfo  {journal} {Phys. Rev. Lett.}\
  }\textbf {\bibinfo {volume} {77}},\ \bibinfo {pages} {3865} (\bibinfo {year}
  {1996})},\ \bibinfo {note} {\textbf{78}, 1396(E) (1997)}\BibitemShut
  {NoStop}%
\bibitem [{\citenamefont {Heyd}\ \emph {et~al.}(2005)\citenamefont {Heyd},
  \citenamefont {Peralta}, \citenamefont {Scuseria},\ and\ \citenamefont
  {Martin}}]{HeydJCP05}%
  \BibitemOpen
  \bibfield  {author} {\bibinfo {author} {\bibfnamefont {J.}~\bibnamefont
  {Heyd}}, \bibinfo {author} {\bibfnamefont {J.~E.}\ \bibnamefont {Peralta}},
  \bibinfo {author} {\bibfnamefont {G.~E.}\ \bibnamefont {Scuseria}}, \ and\
  \bibinfo {author} {\bibfnamefont {R.~L.}\ \bibnamefont {Martin}},\
  }\href@noop {} {\bibfield  {journal} {\bibinfo  {journal} {J. Chem. Phys.}\
  }\textbf {\bibinfo {volume} {123}},\ \bibinfo {pages} {174101} (\bibinfo
  {year} {2005})}\BibitemShut {NoStop}%
\bibitem [{\citenamefont {Aryasetiawan}\ and\ \citenamefont
  {Gunnarsson}(1998)}]{aryasetiawanGWMethod1998}%
  \BibitemOpen
  \bibfield  {author} {\bibinfo {author} {\bibfnamefont {F.}~\bibnamefont
  {Aryasetiawan}}\ and\ \bibinfo {author} {\bibfnamefont {O.}~\bibnamefont
  {Gunnarsson}},\ }\href {\doibase 10.1088/0034-4885/61/3/002} {\bibfield
  {journal} {\bibinfo  {journal} {Rep. Prog. Phys.}\ }\textbf {\bibinfo
  {volume} {61}},\ \bibinfo {pages} {237} (\bibinfo {year} {1998})}\BibitemShut
  {NoStop}%
\bibitem [{\citenamefont {Hedin}(1999)}]{hedinCorrelationEffectsElectron1999}%
  \BibitemOpen
  \bibfield  {author} {\bibinfo {author} {\bibfnamefont {L.}~\bibnamefont
  {Hedin}},\ }\href {\doibase 10.1088/0953-8984/11/42/201} {\bibfield
  {journal} {\bibinfo  {journal} {J. Phys.: Condens. Matter}\ }\textbf
  {\bibinfo {volume} {11}},\ \bibinfo {pages} {R489} (\bibinfo {year}
  {1999})}\BibitemShut {NoStop}%
\bibitem [{\citenamefont {Shishkin}\ \emph {et~al.}(2007)\citenamefont
  {Shishkin}, \citenamefont {Marsman},\ and\ \citenamefont
  {Kresse}}]{ShishkinPRL07}%
  \BibitemOpen
  \bibfield  {author} {\bibinfo {author} {\bibfnamefont {M.}~\bibnamefont
  {Shishkin}}, \bibinfo {author} {\bibfnamefont {M.}~\bibnamefont {Marsman}}, \
  and\ \bibinfo {author} {\bibfnamefont {G.}~\bibnamefont {Kresse}},\
  }\href@noop {} {\bibfield  {journal} {\bibinfo  {journal} {Phys. Rev. Lett.}\
  }\textbf {\bibinfo {volume} {99}},\ \bibinfo {pages} {246403} (\bibinfo
  {year} {2007})}\BibitemShut {NoStop}%
\bibitem [{\citenamefont {Seidl}\ \emph {et~al.}(1996)\citenamefont {Seidl},
  \citenamefont {G\"orling}, \citenamefont {Vogl}, \citenamefont {Majewski},\
  and\ \citenamefont {Levy}}]{seidlGeneralizedKohnShamSchemes1996}%
  \BibitemOpen
  \bibfield  {author} {\bibinfo {author} {\bibfnamefont {A.}~\bibnamefont
  {Seidl}}, \bibinfo {author} {\bibfnamefont {A.}~\bibnamefont {G\"orling}},
  \bibinfo {author} {\bibfnamefont {P.}~\bibnamefont {Vogl}}, \bibinfo {author}
  {\bibfnamefont {J.~A.}\ \bibnamefont {Majewski}}, \ and\ \bibinfo {author}
  {\bibfnamefont {M.}~\bibnamefont {Levy}},\ }\href {\doibase
  10.1103/PhysRevB.53.3764} {\bibfield  {journal} {\bibinfo  {journal}
  {Physical Review B}\ }\textbf {\bibinfo {volume} {53}},\ \bibinfo {pages}
  {3764} (\bibinfo {year} {1996})}\BibitemShut {NoStop}%
\bibitem [{\citenamefont {Becke}(1993)}]{BeckeJCP93b}%
  \BibitemOpen
  \bibfield  {author} {\bibinfo {author} {\bibfnamefont {A.~D.}\ \bibnamefont
  {Becke}},\ }\href@noop {} {\bibfield  {journal} {\bibinfo  {journal} {J.
  Chem. Phys.}\ }\textbf {\bibinfo {volume} {98}},\ \bibinfo {pages} {5648}
  (\bibinfo {year} {1993})}\BibitemShut {NoStop}%
\bibitem [{\citenamefont {Della~Sala}\ \emph {et~al.}(2016)\citenamefont
  {Della~Sala}, \citenamefont {Fabiano},\ and\ \citenamefont
  {Constantin}}]{DellaSalaIJQC16}%
  \BibitemOpen
  \bibfield  {author} {\bibinfo {author} {\bibfnamefont {F.}~\bibnamefont
  {Della~Sala}}, \bibinfo {author} {\bibfnamefont {E.}~\bibnamefont {Fabiano}},
  \ and\ \bibinfo {author} {\bibfnamefont {L.~A.}\ \bibnamefont {Constantin}},\
  }\href@noop {} {\bibfield  {journal} {\bibinfo  {journal} {Int. J. Quantum
  Chem.}\ }\textbf {\bibinfo {volume} {116}},\ \bibinfo {pages} {1641}
  (\bibinfo {year} {2016})}\BibitemShut {NoStop}%
\bibitem [{\citenamefont {Xiao}\ \emph {et~al.}(2013)\citenamefont {Xiao},
  \citenamefont {Sun}, \citenamefont {Ruzsinszky}, \citenamefont {Feng},
  \citenamefont {Haunschild}, \citenamefont {Scuseria},\ and\ \citenamefont
  {Perdew}}]{XiaoPRB13}%
  \BibitemOpen
  \bibfield  {author} {\bibinfo {author} {\bibfnamefont {B.}~\bibnamefont
  {Xiao}}, \bibinfo {author} {\bibfnamefont {J.}~\bibnamefont {Sun}}, \bibinfo
  {author} {\bibfnamefont {A.}~\bibnamefont {Ruzsinszky}}, \bibinfo {author}
  {\bibfnamefont {J.}~\bibnamefont {Feng}}, \bibinfo {author} {\bibfnamefont
  {R.}~\bibnamefont {Haunschild}}, \bibinfo {author} {\bibfnamefont {G.~E.}\
  \bibnamefont {Scuseria}}, \ and\ \bibinfo {author} {\bibfnamefont {J.~P.}\
  \bibnamefont {Perdew}},\ }\href@noop {} {\bibfield  {journal} {\bibinfo
  {journal} {Phys. Rev. B}\ }\textbf {\bibinfo {volume} {88}},\ \bibinfo
  {pages} {184103} (\bibinfo {year} {2013})}\BibitemShut {NoStop}%
\bibitem [{\citenamefont {Yang}\ \emph {et~al.}(2016)\citenamefont {Yang},
  \citenamefont {Peng}, \citenamefont {Sun},\ and\ \citenamefont
  {Perdew}}]{YangPRB16}%
  \BibitemOpen
  \bibfield  {author} {\bibinfo {author} {\bibfnamefont {Z.-h.}\ \bibnamefont
  {Yang}}, \bibinfo {author} {\bibfnamefont {H.}~\bibnamefont {Peng}}, \bibinfo
  {author} {\bibfnamefont {J.}~\bibnamefont {Sun}}, \ and\ \bibinfo {author}
  {\bibfnamefont {J.~P.}\ \bibnamefont {Perdew}},\ }\href@noop {} {\bibfield
  {journal} {\bibinfo  {journal} {Phys. Rev. B}\ }\textbf {\bibinfo {volume}
  {93}},\ \bibinfo {pages} {205205} (\bibinfo {year} {2016})}\BibitemShut
  {NoStop}%
\bibitem [{\citenamefont {Jana}\ \emph {et~al.}(2018)\citenamefont {Jana},
  \citenamefont {Patra},\ and\ \citenamefont {Samal}}]{JanaJCP18}%
  \BibitemOpen
  \bibfield  {author} {\bibinfo {author} {\bibfnamefont {S.}~\bibnamefont
  {Jana}}, \bibinfo {author} {\bibfnamefont {A.}~\bibnamefont {Patra}}, \ and\
  \bibinfo {author} {\bibfnamefont {P.}~\bibnamefont {Samal}},\ }\href@noop {}
  {\bibfield  {journal} {\bibinfo  {journal} {J. Chem. Phys.}\ }\textbf
  {\bibinfo {volume} {149}},\ \bibinfo {pages} {044120} (\bibinfo {year}
  {2018})}\BibitemShut {NoStop}%
\bibitem [{\citenamefont {Armiento}\ and\ \citenamefont
  {K\"{u}mmel}(2013)}]{ArmientoPRL13}%
  \BibitemOpen
  \bibfield  {author} {\bibinfo {author} {\bibfnamefont {R.}~\bibnamefont
  {Armiento}}\ and\ \bibinfo {author} {\bibfnamefont {S.}~\bibnamefont
  {K\"{u}mmel}},\ }\href@noop {} {\bibfield  {journal} {\bibinfo  {journal}
  {Phys. Rev. Lett.}\ }\textbf {\bibinfo {volume} {111}},\ \bibinfo {pages}
  {036402} (\bibinfo {year} {2013})}\BibitemShut {NoStop}%
\bibitem [{\citenamefont {Vl\v{c}ek}\ \emph {et~al.}(2015)\citenamefont
  {Vl\v{c}ek}, \citenamefont {Steinle-Neumann}, \citenamefont {Leppert},
  \citenamefont {Armiento},\ and\ \citenamefont {K\"ummel}}]{VlcekPRB15}%
  \BibitemOpen
  \bibfield  {author} {\bibinfo {author} {\bibfnamefont {V.}~\bibnamefont
  {Vl\v{c}ek}}, \bibinfo {author} {\bibfnamefont {G.}~\bibnamefont
  {Steinle-Neumann}}, \bibinfo {author} {\bibfnamefont {L.}~\bibnamefont
  {Leppert}}, \bibinfo {author} {\bibfnamefont {R.}~\bibnamefont {Armiento}}, \
  and\ \bibinfo {author} {\bibfnamefont {S.}~\bibnamefont {K\"ummel}},\
  }\href@noop {} {\bibfield  {journal} {\bibinfo  {journal} {Phys. Rev. B}\
  }\textbf {\bibinfo {volume} {91}},\ \bibinfo {pages} {035107} (\bibinfo
  {year} {2015})}\BibitemShut {NoStop}%
\bibitem [{\citenamefont {Gritsenko}\ \emph {et~al.}(1995)\citenamefont
  {Gritsenko}, \citenamefont {van Leeuwen}, \citenamefont {van Lenthe},\ and\
  \citenamefont {Baerends}}]{GritsenkoPRA95}%
  \BibitemOpen
  \bibfield  {author} {\bibinfo {author} {\bibfnamefont {O.}~\bibnamefont
  {Gritsenko}}, \bibinfo {author} {\bibfnamefont {R.}~\bibnamefont {van
  Leeuwen}}, \bibinfo {author} {\bibfnamefont {E.}~\bibnamefont {van Lenthe}},
  \ and\ \bibinfo {author} {\bibfnamefont {E.~J.}\ \bibnamefont {Baerends}},\
  }\href@noop {} {\bibfield  {journal} {\bibinfo  {journal} {Phys. Rev. A}\
  }\textbf {\bibinfo {volume} {51}},\ \bibinfo {pages} {1944} (\bibinfo {year}
  {1995})}\BibitemShut {NoStop}%
\bibitem [{\citenamefont {Kuisma}\ \emph {et~al.}(2010)\citenamefont {Kuisma},
  \citenamefont {Ojanen}, \citenamefont {Enkovaara},\ and\ \citenamefont
  {Rantala}}]{KuismaPRB10}%
  \BibitemOpen
  \bibfield  {author} {\bibinfo {author} {\bibfnamefont {M.}~\bibnamefont
  {Kuisma}}, \bibinfo {author} {\bibfnamefont {J.}~\bibnamefont {Ojanen}},
  \bibinfo {author} {\bibfnamefont {J.}~\bibnamefont {Enkovaara}}, \ and\
  \bibinfo {author} {\bibfnamefont {T.~T.}\ \bibnamefont {Rantala}},\
  }\href@noop {} {\bibfield  {journal} {\bibinfo  {journal} {Phys. Rev. B}\
  }\textbf {\bibinfo {volume} {82}},\ \bibinfo {pages} {115106} (\bibinfo
  {year} {2010})}\BibitemShut {NoStop}%
\bibitem [{\citenamefont {Tran}\ and\ \citenamefont
  {Blaha}(2009)}]{tranAccurateBandGaps2009}%
  \BibitemOpen
  \bibfield  {author} {\bibinfo {author} {\bibfnamefont {F.}~\bibnamefont
  {Tran}}\ and\ \bibinfo {author} {\bibfnamefont {P.}~\bibnamefont {Blaha}},\
  }\href {\doibase 10.1103/PhysRevLett.102.226401} {\bibfield  {journal}
  {\bibinfo  {journal} {Phys. Rev. Lett.}\ }\textbf {\bibinfo {volume} {102}},\
  \bibinfo {pages} {226401} (\bibinfo {year} {2009})}\BibitemShut {NoStop}%
\bibitem [{\citenamefont {Ferreira}\ \emph {et~al.}(2008)\citenamefont
  {Ferreira}, \citenamefont {Marques},\ and\ \citenamefont
  {Teles}}]{ferreiraApproximationDensityFunctional2008}%
  \BibitemOpen
  \bibfield  {author} {\bibinfo {author} {\bibfnamefont {L.~G.}\ \bibnamefont
  {Ferreira}}, \bibinfo {author} {\bibfnamefont {M.}~\bibnamefont {Marques}}, \
  and\ \bibinfo {author} {\bibfnamefont {L.~K.}\ \bibnamefont {Teles}},\ }\href
  {\doibase 10.1103/PhysRevB.78.125116} {\bibfield  {journal} {\bibinfo
  {journal} {Phys. Rev. B}\ }\textbf {\bibinfo {volume} {78}},\ \bibinfo
  {pages} {125116} (\bibinfo {year} {2008})}\BibitemShut {NoStop}%
\bibitem [{\citenamefont
  {Slater}(1972)}]{slaterStatisticalExchangeCorrelationSelfConsistent1972}%
  \BibitemOpen
  \bibfield  {author} {\bibinfo {author} {\bibfnamefont {J.~C.}\ \bibnamefont
  {Slater}},\ }in\ \href {\doibase 10.1016/S0065-3276(08)60541-9} {\emph
  {\bibinfo {booktitle} {Advances in {{Quantum Chemistry}}}}},\ Vol.~\bibinfo
  {volume} {6}\ (\bibinfo  {publisher} {{Elsevier}},\ \bibinfo {year} {1972})\
  pp.\ \bibinfo {pages} {1--92}\BibitemShut {NoStop}%
\bibitem [{\citenamefont {Slater}\ and\ \citenamefont
  {Johnson}(1972)}]{slaterSelfConsistentFieldEnsuremathAlpha1972}%
  \BibitemOpen
  \bibfield  {author} {\bibinfo {author} {\bibfnamefont {J.~C.}\ \bibnamefont
  {Slater}}\ and\ \bibinfo {author} {\bibfnamefont {K.~H.}\ \bibnamefont
  {Johnson}},\ }\href {\doibase 10.1103/PhysRevB.5.844} {\bibfield  {journal}
  {\bibinfo  {journal} {Phys. Rev. B}\ }\textbf {\bibinfo {volume} {5}},\
  \bibinfo {pages} {844} (\bibinfo {year} {1972})}\BibitemShut {NoStop}%
\bibitem [{\citenamefont {Ferreira}\ \emph {et~al.}(2013)\citenamefont
  {Ferreira}, \citenamefont {Pel\'a}, \citenamefont {Teles}, \citenamefont
  {Marques}, \citenamefont {Ribeiro~Jr.},\ and\ \citenamefont
  {Furthm\"uller}}]{ferreiraLDA1TechniqueRecent2013}%
  \BibitemOpen
  \bibfield  {author} {\bibinfo {author} {\bibfnamefont {L.~G.}\ \bibnamefont
  {Ferreira}}, \bibinfo {author} {\bibfnamefont {R.~R.}\ \bibnamefont
  {Pel\'a}}, \bibinfo {author} {\bibfnamefont {L.~K.}\ \bibnamefont {Teles}},
  \bibinfo {author} {\bibfnamefont {M.}~\bibnamefont {Marques}}, \bibinfo
  {author} {\bibfnamefont {M.}~\bibnamefont {Ribeiro~Jr.}}, \ and\ \bibinfo
  {author} {\bibfnamefont {J.}~\bibnamefont {Furthm\"uller}},\ }\href {\doibase
  10.1063/1.4848268} {\bibfield  {journal} {\bibinfo  {journal} {AIP Conference
  Proceedings}\ }\textbf {\bibinfo {volume} {1566}},\ \bibinfo {pages} {27}
  (\bibinfo {year} {2013})}\BibitemShut {NoStop}%
\bibitem [{\citenamefont {Ferreira}\ \emph {et~al.}(2011)\citenamefont
  {Ferreira}, \citenamefont {Marques},\ and\ \citenamefont
  {Teles}}]{ferreiraSlaterHalfoccupationTechnique2011}%
  \BibitemOpen
  \bibfield  {author} {\bibinfo {author} {\bibfnamefont {L.~G.}\ \bibnamefont
  {Ferreira}}, \bibinfo {author} {\bibfnamefont {M.}~\bibnamefont {Marques}}, \
  and\ \bibinfo {author} {\bibfnamefont {L.~K.}\ \bibnamefont {Teles}},\ }\href
  {\doibase 10.1063/1.3624562} {\bibfield  {journal} {\bibinfo  {journal} {AIP
  Advances}\ }\textbf {\bibinfo {volume} {1}},\ \bibinfo {pages} {032119}
  (\bibinfo {year} {2011})}\BibitemShut {NoStop}%
\bibitem [{\citenamefont {Pela}\ \emph {et~al.}(2017)\citenamefont {Pela},
  \citenamefont {Gulans},\ and\ \citenamefont
  {Draxl}}]{rodriguespelaLDA1MethodImplemented2017}%
  \BibitemOpen
  \bibfield  {author} {\bibinfo {author} {\bibfnamefont {R.~R.}\ \bibnamefont
  {Pela}}, \bibinfo {author} {\bibfnamefont {A.}~\bibnamefont {Gulans}}, \ and\
  \bibinfo {author} {\bibfnamefont {C.}~\bibnamefont {Draxl}},\ }\href
  {\doibase 10.1016/j.cpc.2017.07.015} {\bibfield  {journal} {\bibinfo
  {journal} {Computer Physics Communications}\ }\textbf {\bibinfo {volume}
  {220}},\ \bibinfo {pages} {263} (\bibinfo {year} {2017})}\BibitemShut
  {NoStop}%
\bibitem [{\citenamefont {Pela}\ \emph {et~al.}(2016)\citenamefont {Pela},
  \citenamefont {Werner}, \citenamefont {Nabok},\ and\ \citenamefont
  {Draxl}}]{rodriguespelaProbingLDA1Method2016}%
  \BibitemOpen
  \bibfield  {author} {\bibinfo {author} {\bibfnamefont {R.~R.}\ \bibnamefont
  {Pela}}, \bibinfo {author} {\bibfnamefont {U.}~\bibnamefont {Werner}},
  \bibinfo {author} {\bibfnamefont {D.}~\bibnamefont {Nabok}}, \ and\ \bibinfo
  {author} {\bibfnamefont {C.}~\bibnamefont {Draxl}},\ }\href {\doibase
  10.1103/PhysRevB.94.235141} {\bibfield  {journal} {\bibinfo  {journal} {Phys.
  Rev. B}\ }\textbf {\bibinfo {volume} {94}},\ \bibinfo {pages} {235141}
  (\bibinfo {year} {2016})}\BibitemShut {NoStop}%
\bibitem [{\citenamefont {Tao}\ \emph {et~al.}(2017)\citenamefont {Tao},
  \citenamefont {Cao},\ and\ \citenamefont
  {Bobbert}}]{taoAccurateEfficientBand2017}%
  \BibitemOpen
  \bibfield  {author} {\bibinfo {author} {\bibfnamefont {S.~X.}\ \bibnamefont
  {Tao}}, \bibinfo {author} {\bibfnamefont {X.}~\bibnamefont {Cao}}, \ and\
  \bibinfo {author} {\bibfnamefont {P.~A.}\ \bibnamefont {Bobbert}},\ }\href
  {\doibase 10.1038/s41598-017-14435-4} {\bibfield  {journal} {\bibinfo
  {journal} {Scientific Reports}\ }\textbf {\bibinfo {volume} {7}},\ \bibinfo
  {pages} {14386} (\bibinfo {year} {2017})}\BibitemShut {NoStop}%
\bibitem [{\citenamefont {Lucatto}\ \emph {et~al.}(2017)\citenamefont
  {Lucatto}, \citenamefont {Assali}, \citenamefont {Pela}, \citenamefont
  {Marques},\ and\ \citenamefont
  {Teles}}]{lucattoGeneralProcedureCalculation2017}%
  \BibitemOpen
  \bibfield  {author} {\bibinfo {author} {\bibfnamefont {B.}~\bibnamefont
  {Lucatto}}, \bibinfo {author} {\bibfnamefont {L.~V.~C.}\ \bibnamefont
  {Assali}}, \bibinfo {author} {\bibfnamefont {R.~R.}\ \bibnamefont {Pela}},
  \bibinfo {author} {\bibfnamefont {M.}~\bibnamefont {Marques}}, \ and\
  \bibinfo {author} {\bibfnamefont {L.~K.}\ \bibnamefont {Teles}},\ }\href
  {\doibase 10.1103/PhysRevB.96.075145} {\bibfield  {journal} {\bibinfo
  {journal} {Phys. Rev. B}\ }\textbf {\bibinfo {volume} {96}},\ \bibinfo
  {pages} {075145} (\bibinfo {year} {2017})}\BibitemShut {NoStop}%
\bibitem [{\citenamefont
  {Ribeiro}(2015{\natexlab{a}})}]{ribeiroApplicationGGA1Excitedstate2015}%
  \BibitemOpen
  \bibfield  {author} {\bibinfo {author} {\bibfnamefont {M.}~\bibnamefont
  {Ribeiro}},\ }\href {\doibase 10.1139/cjp-2014-0381} {\bibfield  {journal}
  {\bibinfo  {journal} {Canadian Journal of Physics}\ }\textbf {\bibinfo
  {volume} {93}},\ \bibinfo {pages} {261} (\bibinfo {year}
  {2015}{\natexlab{a}})}\BibitemShut {NoStop}%
\bibitem [{\citenamefont {Belabbes}\ \emph {et~al.}(2010)\citenamefont
  {Belabbes}, \citenamefont {Zaoui},\ and\ \citenamefont
  {Ferhat}}]{belabbesMagnetismClusteringCrdoped2010}%
  \BibitemOpen
  \bibfield  {author} {\bibinfo {author} {\bibfnamefont {A.}~\bibnamefont
  {Belabbes}}, \bibinfo {author} {\bibfnamefont {A.}~\bibnamefont {Zaoui}}, \
  and\ \bibinfo {author} {\bibfnamefont {M.}~\bibnamefont {Ferhat}},\ }\href
  {\doibase 10.1063/1.3527978} {\bibfield  {journal} {\bibinfo  {journal}
  {Applied Physics Letters}\ }\textbf {\bibinfo {volume} {97}},\ \bibinfo
  {pages} {242509} (\bibinfo {year} {2010})}\BibitemShut {NoStop}%
\bibitem [{\citenamefont {Santos}\ \emph {et~al.}(2012)\citenamefont {Santos},
  \citenamefont {Marques}, \citenamefont {Ferreira}, \citenamefont {Pel\'a},\
  and\ \citenamefont {Teles}}]{santosDigitalMagneticHeterostructures2012}%
  \BibitemOpen
  \bibfield  {author} {\bibinfo {author} {\bibfnamefont {J.~P.~T.}\
  \bibnamefont {Santos}}, \bibinfo {author} {\bibfnamefont {M.}~\bibnamefont
  {Marques}}, \bibinfo {author} {\bibfnamefont {L.~G.}\ \bibnamefont
  {Ferreira}}, \bibinfo {author} {\bibfnamefont {R.~R.}\ \bibnamefont
  {Pel\'a}}, \ and\ \bibinfo {author} {\bibfnamefont {L.~K.}\ \bibnamefont
  {Teles}},\ }\href {\doibase 10.1063/1.4751285} {\bibfield  {journal}
  {\bibinfo  {journal} {Applied Physics Letters}\ }\textbf {\bibinfo {volume}
  {101}},\ \bibinfo {pages} {112403} (\bibinfo {year} {2012})}\BibitemShut
  {NoStop}%
\bibitem [{\citenamefont {Ribeiro}\ \emph {et~al.}(2012)\citenamefont
  {Ribeiro}, \citenamefont {Fonseca}, \citenamefont {Sadowski},\ and\
  \citenamefont {Ramprasad}}]{ribeiroInitioCalculationCdSe2012}%
  \BibitemOpen
  \bibfield  {author} {\bibinfo {author} {\bibfnamefont {M.}~\bibnamefont
  {Ribeiro}}, \bibinfo {author} {\bibfnamefont {L.~R.~C.}\ \bibnamefont
  {Fonseca}}, \bibinfo {author} {\bibfnamefont {T.}~\bibnamefont {Sadowski}}, \
  and\ \bibinfo {author} {\bibfnamefont {R.}~\bibnamefont {Ramprasad}},\ }\href
  {\doibase 10.1063/1.3699054} {\bibfield  {journal} {\bibinfo  {journal}
  {Journal of Applied Physics}\ }\textbf {\bibinfo {volume} {111}},\ \bibinfo
  {pages} {073708} (\bibinfo {year} {2012})}\BibitemShut {NoStop}%
\bibitem [{\citenamefont {Belabbes}\ \emph {et~al.}(2011)\citenamefont
  {Belabbes}, \citenamefont {Furthm\"uller},\ and\ \citenamefont
  {Bechstedt}}]{belabbesElectronicPropertiesPolar2011}%
  \BibitemOpen
  \bibfield  {author} {\bibinfo {author} {\bibfnamefont {A.}~\bibnamefont
  {Belabbes}}, \bibinfo {author} {\bibfnamefont {J.}~\bibnamefont
  {Furthm\"uller}}, \ and\ \bibinfo {author} {\bibfnamefont {F.}~\bibnamefont
  {Bechstedt}},\ }\href {\doibase 10.1103/PhysRevB.84.205304} {\bibfield
  {journal} {\bibinfo  {journal} {Phys. Rev. B}\ }\textbf {\bibinfo {volume}
  {84}},\ \bibinfo {pages} {205304} (\bibinfo {year} {2011})}\BibitemShut
  {NoStop}%
\bibitem [{\citenamefont {K\"ufner}\ \emph {et~al.}(2012)\citenamefont
  {K\"ufner}, \citenamefont {Schleife}, \citenamefont {H\"offling},\ and\
  \citenamefont {Bechstedt}}]{kufnerEnergeticsApproximateQuasiparticle2012a}%
  \BibitemOpen
  \bibfield  {author} {\bibinfo {author} {\bibfnamefont {S.}~\bibnamefont
  {K\"ufner}}, \bibinfo {author} {\bibfnamefont {A.}~\bibnamefont {Schleife}},
  \bibinfo {author} {\bibfnamefont {B.}~\bibnamefont {H\"offling}}, \ and\
  \bibinfo {author} {\bibfnamefont {F.}~\bibnamefont {Bechstedt}},\ }\href
  {\doibase 10.1103/PhysRevB.86.075320} {\bibfield  {journal} {\bibinfo
  {journal} {Phys. Rev. B}\ }\textbf {\bibinfo {volume} {86}},\ \bibinfo
  {pages} {075320} (\bibinfo {year} {2012})}\BibitemShut {NoStop}%
\bibitem [{\citenamefont {Ribeiro}\ \emph {et~al.}(2011)\citenamefont
  {Ribeiro}, \citenamefont {Fonseca},\ and\ \citenamefont
  {Ferreira}}]{ribeiroFirstprinciplesCalculationAlAs2011}%
  \BibitemOpen
  \bibfield  {author} {\bibinfo {author} {\bibfnamefont {M.}~\bibnamefont
  {Ribeiro}}, \bibinfo {author} {\bibfnamefont {L.~R.~C.}\ \bibnamefont
  {Fonseca}}, \ and\ \bibinfo {author} {\bibfnamefont {L.~G.}\ \bibnamefont
  {Ferreira}},\ }\href {\doibase 10.1209/0295-5075/94/27001} {\bibfield
  {journal} {\bibinfo  {journal} {EPL (Europhysics Letters)}\ }\textbf
  {\bibinfo {volume} {94}},\ \bibinfo {pages} {27001} (\bibinfo {year}
  {2011})}\BibitemShut {NoStop}%
\bibitem [{\citenamefont {Ribeiro}\ \emph {et~al.}(2009)\citenamefont
  {Ribeiro}, \citenamefont {Fonseca},\ and\ \citenamefont
  {Ferreira}}]{ribeiroAccuratePredictionSi2009}%
  \BibitemOpen
  \bibfield  {author} {\bibinfo {author} {\bibfnamefont {M.}~\bibnamefont
  {Ribeiro}}, \bibinfo {author} {\bibfnamefont {L.~R.~C.}\ \bibnamefont
  {Fonseca}}, \ and\ \bibinfo {author} {\bibfnamefont {L.~G.}\ \bibnamefont
  {Ferreira}},\ }\href {\doibase 10.1103/PhysRevB.79.241312} {\bibfield
  {journal} {\bibinfo  {journal} {Phys. Rev. B}\ }\textbf {\bibinfo {volume}
  {79}},\ \bibinfo {pages} {241312} (\bibinfo {year} {2009})}\BibitemShut
  {NoStop}%
\bibitem [{\citenamefont {Pela}\ \emph {et~al.}(2015)\citenamefont {Pela},
  \citenamefont {Marques},\ and\ \citenamefont
  {Teles}}]{pelaComparingLDA1HSE032015}%
  \BibitemOpen
  \bibfield  {author} {\bibinfo {author} {\bibfnamefont {R.~R.}\ \bibnamefont
  {Pela}}, \bibinfo {author} {\bibfnamefont {M.}~\bibnamefont {Marques}}, \
  and\ \bibinfo {author} {\bibfnamefont {L.~K.}\ \bibnamefont {Teles}},\ }\href
  {\doibase 10.1088/0953-8984/27/50/505502} {\bibfield  {journal} {\bibinfo
  {journal} {Journal of Physics: Condensed Matter}\ }\textbf {\bibinfo {volume}
  {27}},\ \bibinfo {pages} {505502} (\bibinfo {year} {2015})}\BibitemShut
  {NoStop}%
\bibitem [{\citenamefont {Matusalem}\ \emph {et~al.}(2018)\citenamefont
  {Matusalem}, \citenamefont {Marques}, \citenamefont {Teles}, \citenamefont
  {Filippetti},\ and\ \citenamefont
  {Cappellini}}]{matusalemElectronicPropertiesFluorides2018}%
  \BibitemOpen
  \bibfield  {author} {\bibinfo {author} {\bibfnamefont {F.}~\bibnamefont
  {Matusalem}}, \bibinfo {author} {\bibfnamefont {M.}~\bibnamefont {Marques}},
  \bibinfo {author} {\bibfnamefont {L.~K.}\ \bibnamefont {Teles}}, \bibinfo
  {author} {\bibfnamefont {A.}~\bibnamefont {Filippetti}}, \ and\ \bibinfo
  {author} {\bibfnamefont {G.}~\bibnamefont {Cappellini}},\ }\href {\doibase
  10.1088/1361-648X/aad654} {\bibfield  {journal} {\bibinfo  {journal} {Journal
  of Physics: Condensed Matter}\ }\textbf {\bibinfo {volume} {30}},\ \bibinfo
  {pages} {365501} (\bibinfo {year} {2018})}\BibitemShut {NoStop}%
\bibitem [{\citenamefont {Pel\'a}\ \emph {et~al.}(2012)\citenamefont {Pel\'a},
  \citenamefont {Marques}, \citenamefont {Ferreira}, \citenamefont
  {Furthm\"uller},\ and\ \citenamefont {Teles}}]{pelaGaMnAsPositionMn2012}%
  \BibitemOpen
  \bibfield  {author} {\bibinfo {author} {\bibfnamefont {R.~R.}\ \bibnamefont
  {Pel\'a}}, \bibinfo {author} {\bibfnamefont {M.}~\bibnamefont {Marques}},
  \bibinfo {author} {\bibfnamefont {L.~G.}\ \bibnamefont {Ferreira}}, \bibinfo
  {author} {\bibfnamefont {J.}~\bibnamefont {Furthm\"uller}}, \ and\ \bibinfo
  {author} {\bibfnamefont {L.~K.}\ \bibnamefont {Teles}},\ }\href {\doibase
  10.1063/1.4718602} {\bibfield  {journal} {\bibinfo  {journal} {Applied
  Physics Letters}\ }\textbf {\bibinfo {volume} {100}},\ \bibinfo {pages}
  {202408} (\bibinfo {year} {2012})}\BibitemShut {NoStop}%
\bibitem [{\citenamefont {Pela}\ \emph {et~al.}(2018)\citenamefont {Pela},
  \citenamefont {Gulans},\ and\ \citenamefont
  {Draxl}}]{rodriguespelaLDA1MethodApplied2018}%
  \BibitemOpen
  \bibfield  {author} {\bibinfo {author} {\bibfnamefont {R.~R.}\ \bibnamefont
  {Pela}}, \bibinfo {author} {\bibfnamefont {A.}~\bibnamefont {Gulans}}, \ and\
  \bibinfo {author} {\bibfnamefont {C.}~\bibnamefont {Draxl}},\ }\href
  {\doibase 10.1021/acs.jctc.8b00518} {\bibfield  {journal} {\bibinfo
  {journal} {Journal of Chemical Theory and Computation}\ }\textbf {\bibinfo
  {volume} {14}},\ \bibinfo {pages} {4678} (\bibinfo {year}
  {2018})}\BibitemShut {NoStop}%
\bibitem [{\citenamefont {Xue}\ \emph {et~al.}(2018)\citenamefont {Xue},
  \citenamefont {Yuan}, \citenamefont {Fonseca},\ and\ \citenamefont
  {Miao}}]{xueImprovedLDA1Method2018}%
  \BibitemOpen
  \bibfield  {author} {\bibinfo {author} {\bibfnamefont {K.-H.}\ \bibnamefont
  {Xue}}, \bibinfo {author} {\bibfnamefont {J.-H.}\ \bibnamefont {Yuan}},
  \bibinfo {author} {\bibfnamefont {L.~R.}\ \bibnamefont {Fonseca}}, \ and\
  \bibinfo {author} {\bibfnamefont {X.-S.}\ \bibnamefont {Miao}},\ }\href
  {\doibase 10.1016/j.commatsci.2018.06.036} {\bibfield  {journal} {\bibinfo
  {journal} {Computational Materials Science}\ }\textbf {\bibinfo {volume}
  {153}},\ \bibinfo {pages} {493} (\bibinfo {year} {2018})}\BibitemShut
  {NoStop}%
\bibitem [{\citenamefont {Perdew}\ and\ \citenamefont
  {Levy}(1983)}]{perdewPhysicalContentExact1983}%
  \BibitemOpen
  \bibfield  {author} {\bibinfo {author} {\bibfnamefont {J.~P.}\ \bibnamefont
  {Perdew}}\ and\ \bibinfo {author} {\bibfnamefont {M.}~\bibnamefont {Levy}},\
  }\href {\doibase 10.1103/PhysRevLett.51.1884} {\bibfield  {journal} {\bibinfo
   {journal} {Physical Review Letters}\ }\textbf {\bibinfo {volume} {51}},\
  \bibinfo {pages} {1884} (\bibinfo {year} {1983})}\BibitemShut {NoStop}%
\bibitem [{\citenamefont {Yang}\ \emph {et~al.}(2012)\citenamefont {Yang},
  \citenamefont {Cohen},\ and\ \citenamefont
  {{Mori-S\'anchez}}}]{yangDerivativeDiscontinuityBandgap2012}%
  \BibitemOpen
  \bibfield  {author} {\bibinfo {author} {\bibfnamefont {W.}~\bibnamefont
  {Yang}}, \bibinfo {author} {\bibfnamefont {A.~J.}\ \bibnamefont {Cohen}}, \
  and\ \bibinfo {author} {\bibfnamefont {P.}~\bibnamefont {{Mori-S\'anchez}}},\
  }\href {\doibase 10.1063/1.3702391} {\bibfield  {journal} {\bibinfo
  {journal} {The Journal of Chemical Physics}\ }\textbf {\bibinfo {volume}
  {136}},\ \bibinfo {pages} {204111} (\bibinfo {year} {2012})}\BibitemShut
  {NoStop}%
\bibitem [{\citenamefont {Andrade}\ and\ \citenamefont
  {Aspuru-Guzik}(2011)}]{AndradePRL11}%
  \BibitemOpen
  \bibfield  {author} {\bibinfo {author} {\bibfnamefont {X.}~\bibnamefont
  {Andrade}}\ and\ \bibinfo {author} {\bibfnamefont {A.}~\bibnamefont
  {Aspuru-Guzik}},\ }\href@noop {} {\bibfield  {journal} {\bibinfo  {journal}
  {Phys. Rev. Lett.}\ }\textbf {\bibinfo {volume} {107}},\ \bibinfo {pages}
  {183002} (\bibinfo {year} {2011})}\BibitemShut {NoStop}%
\bibitem [{\citenamefont {Chai}\ and\ \citenamefont {Chen}(2013)}]{ChaiPRL13}%
  \BibitemOpen
  \bibfield  {author} {\bibinfo {author} {\bibfnamefont {J.-D.}\ \bibnamefont
  {Chai}}\ and\ \bibinfo {author} {\bibfnamefont {P.-T.}\ \bibnamefont
  {Chen}},\ }\href@noop {} {\bibfield  {journal} {\bibinfo  {journal} {Phys.
  Rev. Lett.}\ }\textbf {\bibinfo {volume} {110}},\ \bibinfo {pages} {033002}
  (\bibinfo {year} {2013})}\BibitemShut {NoStop}%
\bibitem [{\citenamefont {Kraisler}\ and\ \citenamefont
  {Kronik}(2014)}]{KraislerJCP14}%
  \BibitemOpen
  \bibfield  {author} {\bibinfo {author} {\bibfnamefont {E.}~\bibnamefont
  {Kraisler}}\ and\ \bibinfo {author} {\bibfnamefont {L.}~\bibnamefont
  {Kronik}},\ }\href@noop {} {\bibfield  {journal} {\bibinfo  {journal} {J.
  Chem. Phys.}\ }\textbf {\bibinfo {volume} {140}},\ \bibinfo {pages} {18A540}
  (\bibinfo {year} {2014})}\BibitemShut {NoStop}%
\bibitem [{\citenamefont {Gr\"{u}ning}\ \emph {et~al.}(2006)\citenamefont
  {Gr\"{u}ning}, \citenamefont {Marini},\ and\ \citenamefont
  {Rubio}}]{GruningJCP06}%
  \BibitemOpen
  \bibfield  {author} {\bibinfo {author} {\bibfnamefont {M.}~\bibnamefont
  {Gr\"{u}ning}}, \bibinfo {author} {\bibfnamefont {A.}~\bibnamefont {Marini}},
  \ and\ \bibinfo {author} {\bibfnamefont {A.}~\bibnamefont {Rubio}},\
  }\href@noop {} {\bibfield  {journal} {\bibinfo  {journal} {J. Chem. Phys.}\
  }\textbf {\bibinfo {volume} {124}},\ \bibinfo {pages} {154108} (\bibinfo
  {year} {2006})}\BibitemShut {NoStop}%
\bibitem [{\citenamefont {Gr\"uning}\ \emph {et~al.}(2006)\citenamefont
  {Gr\"uning}, \citenamefont {Marini},\ and\ \citenamefont
  {Rubio}}]{GruningPRB06}%
  \BibitemOpen
  \bibfield  {author} {\bibinfo {author} {\bibfnamefont {M.}~\bibnamefont
  {Gr\"uning}}, \bibinfo {author} {\bibfnamefont {A.}~\bibnamefont {Marini}}, \
  and\ \bibinfo {author} {\bibfnamefont {A.}~\bibnamefont {Rubio}},\
  }\href@noop {} {\bibfield  {journal} {\bibinfo  {journal} {Phys. Rev. B}\
  }\textbf {\bibinfo {volume} {74}},\ \bibinfo {pages} {161103(R)} (\bibinfo
  {year} {2006})}\BibitemShut {NoStop}%
\bibitem [{\citenamefont {Janak}(1978)}]{janakProofThatFrac1978}%
  \BibitemOpen
  \bibfield  {author} {\bibinfo {author} {\bibfnamefont {J.~F.}\ \bibnamefont
  {Janak}},\ }\href {\doibase 10.1103/PhysRevB.18.7165} {\bibfield  {journal}
  {\bibinfo  {journal} {Phys. Rev. B}\ }\textbf {\bibinfo {volume} {18}},\
  \bibinfo {pages} {7165} (\bibinfo {year} {1978})}\BibitemShut {NoStop}%
\bibitem [{\citenamefont {K{\"u}mmel}\ and\ \citenamefont
  {Kronik}(2008)}]{KuemmelRMP08}%
  \BibitemOpen
  \bibfield  {author} {\bibinfo {author} {\bibfnamefont {S.}~\bibnamefont
  {K{\"u}mmel}}\ and\ \bibinfo {author} {\bibfnamefont {L.}~\bibnamefont
  {Kronik}},\ }\href@noop {} {\bibfield  {journal} {\bibinfo  {journal} {Rev.
  Mod. Phys.}\ }\textbf {\bibinfo {volume} {80}},\ \bibinfo {pages} {3}
  (\bibinfo {year} {2008})}\BibitemShut {NoStop}%
\bibitem [{\citenamefont {Ataide}\ \emph {et~al.}(2017)\citenamefont {Ataide},
  \citenamefont {Pel\'a}, \citenamefont {Marques}, \citenamefont {Teles},
  \citenamefont {Furthm\"uller},\ and\ \citenamefont
  {Bechstedt}}]{ataideFastAccurateApproximate2017}%
  \BibitemOpen
  \bibfield  {author} {\bibinfo {author} {\bibfnamefont {C.~A.}\ \bibnamefont
  {Ataide}}, \bibinfo {author} {\bibfnamefont {R.~R.}\ \bibnamefont {Pel\'a}},
  \bibinfo {author} {\bibfnamefont {M.}~\bibnamefont {Marques}}, \bibinfo
  {author} {\bibfnamefont {L.~K.}\ \bibnamefont {Teles}}, \bibinfo {author}
  {\bibfnamefont {J.}~\bibnamefont {Furthm\"uller}}, \ and\ \bibinfo {author}
  {\bibfnamefont {F.}~\bibnamefont {Bechstedt}},\ }\href {\doibase
  10.1103/PhysRevB.95.045126} {\bibfield  {journal} {\bibinfo  {journal} {Phys.
  Rev. B}\ }\textbf {\bibinfo {volume} {95}},\ \bibinfo {pages} {045126}
  (\bibinfo {year} {2017})}\BibitemShut {NoStop}%
\bibitem [{\citenamefont
  {Ribeiro}(2015{\natexlab{b}})}]{ribeiroInitioQuasiparticleApproximation2015}%
  \BibitemOpen
  \bibfield  {author} {\bibinfo {author} {\bibfnamefont {M.}~\bibnamefont
  {Ribeiro}},\ }\href {\doibase 10.1063/1.4922337} {\bibfield  {journal}
  {\bibinfo  {journal} {Journal of Applied Physics}\ }\textbf {\bibinfo
  {volume} {117}},\ \bibinfo {pages} {234302} (\bibinfo {year}
  {2015}{\natexlab{b}})}\BibitemShut {NoStop}%
\bibitem [{\citenamefont {Blaha}\ \emph {et~al.}(2018)\citenamefont {Blaha},
  \citenamefont {Schwarz}, \citenamefont {Madsen}, \citenamefont {Kvasnicka},
  \citenamefont {Luitz}, \citenamefont {Laskowski}, \citenamefont {Tran},\ and\
  \citenamefont {Marks}}]{wien2k}%
  \BibitemOpen
  \bibfield  {author} {\bibinfo {author} {\bibfnamefont {P.}~\bibnamefont
  {Blaha}}, \bibinfo {author} {\bibfnamefont {K.}~\bibnamefont {Schwarz}},
  \bibinfo {author} {\bibfnamefont {G.~K.~H.}\ \bibnamefont {Madsen}}, \bibinfo
  {author} {\bibfnamefont {D.}~\bibnamefont {Kvasnicka}}, \bibinfo {author}
  {\bibfnamefont {J.}~\bibnamefont {Luitz}}, \bibinfo {author} {\bibfnamefont
  {R.}~\bibnamefont {Laskowski}}, \bibinfo {author} {\bibfnamefont
  {F.}~\bibnamefont {Tran}}, \ and\ \bibinfo {author} {\bibfnamefont {L.~D.}\
  \bibnamefont {Marks}},\ }\href@noop {} {\emph {\bibinfo {title} {{{WIEN2k}}:
  {{An Augmented Plane Wave}} plus {{Local Orbitals Program}} for {{Calculating
  Crystal Properties}}}}}\ (\bibinfo  {publisher} {{Vienna University of
  Technology, Austria}},\ \bibinfo {year} {2018})\BibitemShut {NoStop}%
\bibitem [{\citenamefont {Andersen}(1975)}]{AndersenPRB75}%
  \BibitemOpen
  \bibfield  {author} {\bibinfo {author} {\bibfnamefont {O.~K.}\ \bibnamefont
  {Andersen}},\ }\href@noop {} {\bibfield  {journal} {\bibinfo  {journal}
  {Phys. Rev. B}\ }\textbf {\bibinfo {volume} {12}},\ \bibinfo {pages} {3060}
  (\bibinfo {year} {1975})}\BibitemShut {NoStop}%
\bibitem [{\citenamefont {Singh}\ and\ \citenamefont
  {Nordstr{\"{o}}m}(2006)}]{Singh}%
  \BibitemOpen
  \bibfield  {author} {\bibinfo {author} {\bibfnamefont {D.~J.}\ \bibnamefont
  {Singh}}\ and\ \bibinfo {author} {\bibfnamefont {L.}~\bibnamefont
  {Nordstr{\"{o}}m}},\ }\href@noop {} {\emph {\bibinfo {title} {Planewaves,
  Pseudopotentials, and the LAPW Method, 2nd ed.}}}\ (\bibinfo  {publisher}
  {Springer},\ \bibinfo {address} {New York},\ \bibinfo {year}
  {2006})\BibitemShut {NoStop}%
\bibitem [{SM_()}]{SM_LDA-half}%
  \BibitemOpen
  \href@noop {} {}\bibinfo {howpublished} {See Supplemental Material at
  http://link.aps.org/supplemental/ for the experimental lattice constants of
  the solids used for the calculations.}\BibitemShut {Stop}%
\bibitem [{\citenamefont {Tran}\ and\ \citenamefont
  {Blaha}(2017)}]{TranJPCA17}%
  \BibitemOpen
  \bibfield  {author} {\bibinfo {author} {\bibfnamefont {F.}~\bibnamefont
  {Tran}}\ and\ \bibinfo {author} {\bibfnamefont {P.}~\bibnamefont {Blaha}},\
  }\href@noop {} {\bibfield  {journal} {\bibinfo  {journal} {J. Phys. Chem. A}\
  }\textbf {\bibinfo {volume} {121}},\ \bibinfo {pages} {3318} (\bibinfo {year}
  {2017})}\BibitemShut {NoStop}%
\bibitem [{\citenamefont {Tran}\ \emph {et~al.}(2018)\citenamefont {Tran},
  \citenamefont {Ehsan},\ and\ \citenamefont {Blaha}}]{TranPRM18}%
  \BibitemOpen
  \bibfield  {author} {\bibinfo {author} {\bibfnamefont {F.}~\bibnamefont
  {Tran}}, \bibinfo {author} {\bibfnamefont {S.}~\bibnamefont {Ehsan}}, \ and\
  \bibinfo {author} {\bibfnamefont {P.}~\bibnamefont {Blaha}},\ }\href@noop {}
  {\bibfield  {journal} {\bibinfo  {journal} {Phys. Rev. Materials}\ }\textbf
  {\bibinfo {volume} {2}},\ \bibinfo {pages} {023802} (\bibinfo {year}
  {2018})}\BibitemShut {NoStop}%
\bibitem [{\citenamefont {Lee}\ \emph {et~al.}(2016{\natexlab{a}})\citenamefont
  {Lee}, \citenamefont {Seko}, \citenamefont {Shitara}, \citenamefont
  {Nakayama},\ and\ \citenamefont {Tanaka}}]{LeePRB16}%
  \BibitemOpen
  \bibfield  {author} {\bibinfo {author} {\bibfnamefont {J.}~\bibnamefont
  {Lee}}, \bibinfo {author} {\bibfnamefont {A.}~\bibnamefont {Seko}}, \bibinfo
  {author} {\bibfnamefont {K.}~\bibnamefont {Shitara}}, \bibinfo {author}
  {\bibfnamefont {K.}~\bibnamefont {Nakayama}}, \ and\ \bibinfo {author}
  {\bibfnamefont {I.}~\bibnamefont {Tanaka}},\ }\href@noop {} {\bibfield
  {journal} {\bibinfo  {journal} {Phys. Rev. B}\ }\textbf {\bibinfo {volume}
  {93}},\ \bibinfo {pages} {115104} (\bibinfo {year}
  {2016}{\natexlab{a}})}\BibitemShut {NoStop}%
\bibitem [{\citenamefont {Nakano}\ and\ \citenamefont
  {Sakai}(2018)}]{NakanoJAP18}%
  \BibitemOpen
  \bibfield  {author} {\bibinfo {author} {\bibfnamefont {K.}~\bibnamefont
  {Nakano}}\ and\ \bibinfo {author} {\bibfnamefont {T.}~\bibnamefont {Sakai}},\
  }\href@noop {} {\bibfield  {journal} {\bibinfo  {journal} {J. Appl. Phys.}\
  }\textbf {\bibinfo {volume} {123}},\ \bibinfo {pages} {015104} (\bibinfo
  {year} {2018})}\BibitemShut {NoStop}%
\bibitem [{\citenamefont {Lucero}\ \emph {et~al.}(2012)\citenamefont {Lucero},
  \citenamefont {Henderson},\ and\ \citenamefont
  {Scuseria}}]{luceroImprovedSemiconductorLattice2012}%
  \BibitemOpen
  \bibfield  {author} {\bibinfo {author} {\bibfnamefont {M.~J.}\ \bibnamefont
  {Lucero}}, \bibinfo {author} {\bibfnamefont {T.~M.}\ \bibnamefont
  {Henderson}}, \ and\ \bibinfo {author} {\bibfnamefont {G.~E.}\ \bibnamefont
  {Scuseria}},\ }\href {\doibase 10.1088/0953-8984/24/14/145504} {\bibfield
  {journal} {\bibinfo  {journal} {Journal of Physics: Condensed Matter}\
  }\textbf {\bibinfo {volume} {24}},\ \bibinfo {pages} {145504} (\bibinfo
  {year} {2012})}\BibitemShut {NoStop}%
\bibitem [{\citenamefont {Crowley}\ \emph {et~al.}(2016)\citenamefont
  {Crowley}, \citenamefont {{Tahir-Kheli}},\ and\ \citenamefont
  {Goddard}}]{crowleyResolutionBandGap2016a}%
  \BibitemOpen
  \bibfield  {author} {\bibinfo {author} {\bibfnamefont {J.~M.}\ \bibnamefont
  {Crowley}}, \bibinfo {author} {\bibfnamefont {J.}~\bibnamefont
  {{Tahir-Kheli}}}, \ and\ \bibinfo {author} {\bibfnamefont {W.~A.}\
  \bibnamefont {Goddard}},\ }\href {\doibase 10.1021/acs.jpclett.5b02870}
  {\bibfield  {journal} {\bibinfo  {journal} {The Journal of Physical Chemistry
  Letters}\ }\textbf {\bibinfo {volume} {7}},\ \bibinfo {pages} {1198}
  (\bibinfo {year} {2016})}\BibitemShut {NoStop}%
\bibitem [{\citenamefont {Bagheri}\ and\ \citenamefont
  {Blaha}(2019)}]{bagheriDFTCalculationsEnergy2019}%
  \BibitemOpen
  \bibfield  {author} {\bibinfo {author} {\bibfnamefont {M.}~\bibnamefont
  {Bagheri}}\ and\ \bibinfo {author} {\bibfnamefont {P.}~\bibnamefont
  {Blaha}},\ }\href {\doibase 10.1016/j.elspec.2018.11.002} {\bibfield
  {journal} {\bibinfo  {journal} {Journal of Electron Spectroscopy and Related
  Phenomena}\ }\textbf {\bibinfo {volume} {230}},\ \bibinfo {pages} {1}
  (\bibinfo {year} {2019})}\BibitemShut {NoStop}%
\bibitem [{\citenamefont {Lee}\ \emph {et~al.}(2016{\natexlab{b}})\citenamefont
  {Lee}, \citenamefont {Seko}, \citenamefont {Shitara}, \citenamefont
  {Nakayama},\ and\ \citenamefont {Tanaka}}]{leePredictionModelBand2016}%
  \BibitemOpen
  \bibfield  {author} {\bibinfo {author} {\bibfnamefont {J.}~\bibnamefont
  {Lee}}, \bibinfo {author} {\bibfnamefont {A.}~\bibnamefont {Seko}}, \bibinfo
  {author} {\bibfnamefont {K.}~\bibnamefont {Shitara}}, \bibinfo {author}
  {\bibfnamefont {K.}~\bibnamefont {Nakayama}}, \ and\ \bibinfo {author}
  {\bibfnamefont {I.}~\bibnamefont {Tanaka}},\ }\href {\doibase
  10.1103/PhysRevB.93.115104} {\bibfield  {journal} {\bibinfo  {journal}
  {Physical Review B}\ }\textbf {\bibinfo {volume} {93}},\ \bibinfo {pages}
  {12} (\bibinfo {year} {2016}{\natexlab{b}})}\BibitemShut {NoStop}%
\bibitem [{\citenamefont {Yim}\ \emph {et~al.}(1972)\citenamefont {Yim},
  \citenamefont {Dismukes}, \citenamefont {Stofko},\ and\ \citenamefont
  {Paff}}]{yimSynthesisPropertiesBeTe1972}%
  \BibitemOpen
  \bibfield  {author} {\bibinfo {author} {\bibfnamefont {W.}~\bibnamefont
  {Yim}}, \bibinfo {author} {\bibfnamefont {J.}~\bibnamefont {Dismukes}},
  \bibinfo {author} {\bibfnamefont {E.}~\bibnamefont {Stofko}}, \ and\ \bibinfo
  {author} {\bibfnamefont {R.}~\bibnamefont {Paff}},\ }\href {\doibase
  10.1016/0022-3697(72)90032-7} {\bibfield  {journal} {\bibinfo  {journal}
  {Journal of Physics and Chemistry of Solids}\ }\textbf {\bibinfo {volume}
  {33}},\ \bibinfo {pages} {501} (\bibinfo {year} {1972})}\BibitemShut
  {NoStop}%
\bibitem [{\citenamefont {Nagelstra\ss{}er}\ \emph {et~al.}(1998)\citenamefont
  {Nagelstra\ss{}er}, \citenamefont {Dr\"oge}, \citenamefont {Steinr\"uck},
  \citenamefont {Fischer}, \citenamefont {Litz}, \citenamefont {Waag},
  \citenamefont {Landwehr}, \citenamefont {Fleszar},\ and\ \citenamefont
  {Hanke}}]{nagelstrasserBandStructureBeTe1998}%
  \BibitemOpen
  \bibfield  {author} {\bibinfo {author} {\bibfnamefont {M.}~\bibnamefont
  {Nagelstra\ss{}er}}, \bibinfo {author} {\bibfnamefont {H.}~\bibnamefont
  {Dr\"oge}}, \bibinfo {author} {\bibfnamefont {H.-P.}\ \bibnamefont
  {Steinr\"uck}}, \bibinfo {author} {\bibfnamefont {F.}~\bibnamefont
  {Fischer}}, \bibinfo {author} {\bibfnamefont {T.}~\bibnamefont {Litz}},
  \bibinfo {author} {\bibfnamefont {A.}~\bibnamefont {Waag}}, \bibinfo {author}
  {\bibfnamefont {G.}~\bibnamefont {Landwehr}}, \bibinfo {author}
  {\bibfnamefont {A.}~\bibnamefont {Fleszar}}, \ and\ \bibinfo {author}
  {\bibfnamefont {W.}~\bibnamefont {Hanke}},\ }\href {\doibase
  10.1103/PhysRevB.58.10394} {\bibfield  {journal} {\bibinfo  {journal}
  {Physical Review B}\ }\textbf {\bibinfo {volume} {58}},\ \bibinfo {pages}
  {10394} (\bibinfo {year} {1998})}\BibitemShut {NoStop}%
\bibitem [{\citenamefont {Hummelsh\o{}j}\ \emph {et~al.}(2010)\citenamefont
  {Hummelsh\o{}j}, \citenamefont {Blomqvist}, \citenamefont {Datta},
  \citenamefont {Vegge}, \citenamefont {Rossmeisl}, \citenamefont {Thygesen},
  \citenamefont {Luntz}, \citenamefont {Jacobsen},\ and\ \citenamefont
  {N\o{}rskov}}]{hummelshojCommunicationsElementaryOxygen2010}%
  \BibitemOpen
  \bibfield  {author} {\bibinfo {author} {\bibfnamefont {J.~S.}\ \bibnamefont
  {Hummelsh\o{}j}}, \bibinfo {author} {\bibfnamefont {J.}~\bibnamefont
  {Blomqvist}}, \bibinfo {author} {\bibfnamefont {S.}~\bibnamefont {Datta}},
  \bibinfo {author} {\bibfnamefont {T.}~\bibnamefont {Vegge}}, \bibinfo
  {author} {\bibfnamefont {J.}~\bibnamefont {Rossmeisl}}, \bibinfo {author}
  {\bibfnamefont {K.~S.}\ \bibnamefont {Thygesen}}, \bibinfo {author}
  {\bibfnamefont {A.~C.}\ \bibnamefont {Luntz}}, \bibinfo {author}
  {\bibfnamefont {K.~W.}\ \bibnamefont {Jacobsen}}, \ and\ \bibinfo {author}
  {\bibfnamefont {J.~K.}\ \bibnamefont {N\o{}rskov}},\ }\href {\doibase
  10.1063/1.3298994} {\bibfield  {journal} {\bibinfo  {journal} {The Journal of
  Chemical Physics}\ }\textbf {\bibinfo {volume} {132}},\ \bibinfo {pages}
  {071101} (\bibinfo {year} {2010})}\BibitemShut {NoStop}%
\bibitem [{\citenamefont {Gillen}\ and\ \citenamefont
  {Robertson}(2013)}]{gillenAccurateScreenedExchange2013}%
  \BibitemOpen
  \bibfield  {author} {\bibinfo {author} {\bibfnamefont {R.}~\bibnamefont
  {Gillen}}\ and\ \bibinfo {author} {\bibfnamefont {J.}~\bibnamefont
  {Robertson}},\ }\href {\doibase 10.1088/0953-8984/25/16/165502} {\bibfield
  {journal} {\bibinfo  {journal} {Journal of Physics: Condensed Matter}\
  }\textbf {\bibinfo {volume} {25}},\ \bibinfo {pages} {165502} (\bibinfo
  {year} {2013})}\BibitemShut {NoStop}%
\bibitem [{\citenamefont {Wang}\ \emph {et~al.}(2016)\citenamefont {Wang},
  \citenamefont {Lany}, \citenamefont {Ghanbaja}, \citenamefont
  {{Fagot-Revurat}}, \citenamefont {Chen}, \citenamefont {Soldera},
  \citenamefont {Horwat}, \citenamefont {M\"ucklich},\ and\ \citenamefont
  {Pierson}}]{wangElectronicStructuresMathrmC2016}%
  \BibitemOpen
  \bibfield  {author} {\bibinfo {author} {\bibfnamefont {Y.}~\bibnamefont
  {Wang}}, \bibinfo {author} {\bibfnamefont {S.}~\bibnamefont {Lany}}, \bibinfo
  {author} {\bibfnamefont {J.}~\bibnamefont {Ghanbaja}}, \bibinfo {author}
  {\bibfnamefont {Y.}~\bibnamefont {{Fagot-Revurat}}}, \bibinfo {author}
  {\bibfnamefont {Y.~P.}\ \bibnamefont {Chen}}, \bibinfo {author}
  {\bibfnamefont {F.}~\bibnamefont {Soldera}}, \bibinfo {author} {\bibfnamefont
  {D.}~\bibnamefont {Horwat}}, \bibinfo {author} {\bibfnamefont
  {F.}~\bibnamefont {M\"ucklich}}, \ and\ \bibinfo {author} {\bibfnamefont
  {J.~F.}\ \bibnamefont {Pierson}},\ }\href {\doibase
  10.1103/PhysRevB.94.245418} {\bibfield  {journal} {\bibinfo  {journal} {Phys.
  Rev. B}\ }\textbf {\bibinfo {volume} {94}},\ \bibinfo {pages} {245418}
  (\bibinfo {year} {2016})}\BibitemShut {NoStop}%
\bibitem [{\citenamefont {Terakura}\ \emph {et~al.}(1984)\citenamefont
  {Terakura}, \citenamefont {Oguchi}, \citenamefont {Williams},\ and\
  \citenamefont {K\"{u}bler}}]{TerakuraPRB84}%
  \BibitemOpen
  \bibfield  {author} {\bibinfo {author} {\bibfnamefont {K.}~\bibnamefont
  {Terakura}}, \bibinfo {author} {\bibfnamefont {T.}~\bibnamefont {Oguchi}},
  \bibinfo {author} {\bibfnamefont {A.~R.}\ \bibnamefont {Williams}}, \ and\
  \bibinfo {author} {\bibfnamefont {J.}~\bibnamefont {K\"{u}bler}},\
  }\href@noop {} {\bibfield  {journal} {\bibinfo  {journal} {Phys. Rev. B}\
  }\textbf {\bibinfo {volume} {30}},\ \bibinfo {pages} {4734} (\bibinfo {year}
  {1984})}\BibitemShut {NoStop}%
\bibitem [{\citenamefont {Anisimov}\ \emph {et~al.}(1991)\citenamefont
  {Anisimov}, \citenamefont {Zaanen},\ and\ \citenamefont
  {Andersen}}]{AnisimovPRB91}%
  \BibitemOpen
  \bibfield  {author} {\bibinfo {author} {\bibfnamefont {V.~I.}\ \bibnamefont
  {Anisimov}}, \bibinfo {author} {\bibfnamefont {J.}~\bibnamefont {Zaanen}}, \
  and\ \bibinfo {author} {\bibfnamefont {O.~K.}\ \bibnamefont {Andersen}},\
  }\href@noop {} {\bibfield  {journal} {\bibinfo  {journal} {Phys. Rev. B}\
  }\textbf {\bibinfo {volume} {44}},\ \bibinfo {pages} {943} (\bibinfo {year}
  {1991})}\BibitemShut {NoStop}%
\bibitem [{\citenamefont {Tran}\ \emph {et~al.}(2006)\citenamefont {Tran},
  \citenamefont {Blaha}, \citenamefont {Schwarz},\ and\ \citenamefont
  {Nov\'{a}k}}]{TranPRB06}%
  \BibitemOpen
  \bibfield  {author} {\bibinfo {author} {\bibfnamefont {F.}~\bibnamefont
  {Tran}}, \bibinfo {author} {\bibfnamefont {P.}~\bibnamefont {Blaha}},
  \bibinfo {author} {\bibfnamefont {K.}~\bibnamefont {Schwarz}}, \ and\
  \bibinfo {author} {\bibfnamefont {P.}~\bibnamefont {Nov\'{a}k}},\ }\href@noop
  {} {\bibfield  {journal} {\bibinfo  {journal} {Phys. Rev. B}\ }\textbf
  {\bibinfo {volume} {74}},\ \bibinfo {pages} {155108} (\bibinfo {year}
  {2006})}\BibitemShut {NoStop}%
\bibitem [{\citenamefont {Marsman}\ \emph {et~al.}(2008)\citenamefont
  {Marsman}, \citenamefont {Paier}, \citenamefont {Stroppa},\ and\
  \citenamefont {Kresse}}]{MarsmanJPCM08}%
  \BibitemOpen
  \bibfield  {author} {\bibinfo {author} {\bibfnamefont {M.}~\bibnamefont
  {Marsman}}, \bibinfo {author} {\bibfnamefont {J.}~\bibnamefont {Paier}},
  \bibinfo {author} {\bibfnamefont {A.}~\bibnamefont {Stroppa}}, \ and\
  \bibinfo {author} {\bibfnamefont {G.}~\bibnamefont {Kresse}},\ }\href@noop {}
  {\bibfield  {journal} {\bibinfo  {journal} {J. Phys.: Condens. Matter}\
  }\textbf {\bibinfo {volume} {20}},\ \bibinfo {pages} {064201} (\bibinfo
  {year} {2008})}\BibitemShut {NoStop}%
\bibitem [{\citenamefont {Gerosa}\ \emph {et~al.}(2018)\citenamefont {Gerosa},
  \citenamefont {Bottani}, \citenamefont {Valentin}, \citenamefont {Onida},\
  and\ \citenamefont
  {Pacchioni}}]{gerosaAccuracyDielectricdependentHybrid2018}%
  \BibitemOpen
  \bibfield  {author} {\bibinfo {author} {\bibfnamefont {M.}~\bibnamefont
  {Gerosa}}, \bibinfo {author} {\bibfnamefont {C.~E.}\ \bibnamefont {Bottani}},
  \bibinfo {author} {\bibfnamefont {C.~D.}\ \bibnamefont {Valentin}}, \bibinfo
  {author} {\bibfnamefont {G.}~\bibnamefont {Onida}}, \ and\ \bibinfo {author}
  {\bibfnamefont {G.}~\bibnamefont {Pacchioni}},\ }\href {\doibase
  10.1088/1361-648X/aa9725} {\bibfield  {journal} {\bibinfo  {journal} {J.
  Phys.: Condens. Matter}\ }\textbf {\bibinfo {volume} {30}},\ \bibinfo {pages}
  {044003} (\bibinfo {year} {2018})}\BibitemShut {NoStop}%
\bibitem [{\citenamefont {Koller}\ \emph {et~al.}(2011)\citenamefont {Koller},
  \citenamefont {Tran},\ and\ \citenamefont
  {Blaha}}]{kollerMeritsLimitsModified2011}%
  \BibitemOpen
  \bibfield  {author} {\bibinfo {author} {\bibfnamefont {D.}~\bibnamefont
  {Koller}}, \bibinfo {author} {\bibfnamefont {F.}~\bibnamefont {Tran}}, \ and\
  \bibinfo {author} {\bibfnamefont {P.}~\bibnamefont {Blaha}},\ }\href
  {\doibase 10.1103/PhysRevB.83.195134} {\bibfield  {journal} {\bibinfo
  {journal} {Phys. Rev. B}\ }\textbf {\bibinfo {volume} {83}},\ \bibinfo
  {pages} {195134} (\bibinfo {year} {2011})}\BibitemShut {NoStop}%
\bibitem [{\citenamefont {Matusalem}\ \emph {et~al.}(2013)\citenamefont
  {Matusalem}, \citenamefont {Ribeiro}, \citenamefont {Marques}, \citenamefont
  {Pel\'a}, \citenamefont {Ferreira},\ and\ \citenamefont
  {Teles}}]{matusalemCombinedLDALDA12013}%
  \BibitemOpen
  \bibfield  {author} {\bibinfo {author} {\bibfnamefont {F.}~\bibnamefont
  {Matusalem}}, \bibinfo {author} {\bibfnamefont {J.}~\bibnamefont {Ribeiro},
  \bibfnamefont {Mauro}}, \bibinfo {author} {\bibfnamefont {M.}~\bibnamefont
  {Marques}}, \bibinfo {author} {\bibfnamefont {R.~R.}\ \bibnamefont {Pel\'a}},
  \bibinfo {author} {\bibfnamefont {L.~G.}\ \bibnamefont {Ferreira}}, \ and\
  \bibinfo {author} {\bibfnamefont {L.~K.}\ \bibnamefont {Teles}},\ }\href
  {\doibase 10.1103/PhysRevB.88.224102} {\bibfield  {journal} {\bibinfo
  {journal} {Phys. Rev. B}\ }\textbf {\bibinfo {volume} {88}},\ \bibinfo
  {pages} {224102} (\bibinfo {year} {2013})}\BibitemShut {NoStop}%
\end{thebibliography}%

\end{document}